%% file: bare_jrnl_new_sample4.tex
\documentclass[lettersize,journal]{IEEEtran}
\usepackage{amsmath,amsfonts}
\usepackage{algorithmic}
\usepackage{algorithm}
\usepackage{array}
\usepackage[caption=false,font=normalsize,labelfont=sf,textfont=sf]{subfig}
\usepackage{textcomp}
\usepackage{stfloats}
\usepackage{url}
\usepackage{verbatim}
\usepackage{graphicx}
\usepackage{cite}
\usepackage{textcomp}
\usepackage{stfloats}
\usepackage{url}
\usepackage{verbatim}
\usepackage{times}                     

\usepackage{tabu}                      
\usepackage{multirow}
\usepackage{mathptmx}                  
\usepackage{glossaries}

\usepackage{placeins}
\usepackage{adjustbox}
\usepackage{mathptmx}                  
\usepackage{glossaries}
\usepackage{graphicx}
\usepackage{xspace}                    
\usepackage{svg}
\usepackage{verbatim}

\usepackage{placeins}
\usepackage{adjustbox}

\usepackage{enumitem}
\usepackage{rotating}
\usepackage{longtable}
\usepackage{makecell} %
\usepackage{xcolor}

\usepackage[colorlinks=true,
            citecolor=blue,
            linkcolor=black,
            urlcolor=blue]{hyperref}

\usepackage{cite}

 \usepackage{orcidlink}
\usepackage[capitalise]{cleveref}
\input{commands}

\makeglossaries

\input{acronyms}

\usepackage{multirow}
\usepackage{booktabs}

\begin{document}
\title{AIvaluateXR: An Evaluation Framework for on-Device AI in XR with Benchmarking Results}
\author{ Dawar~Khan, Xinyu Liu, Omar Mena, Donggang Jia, Alexandre Kouyoumdjian, Ivan~Viola
\thanks{Manuscript received April 19, 20XX; revised August 16, 20XX. 
This research has been funded by KAUST Competitive Research Grants ORFS-CRG12-2024-6422.}
\thanks{ Dawar~Khan,  Xinyu Liu, Omar Mena, Donggang Jia, Alexandre Kouyoumdjian, Ivan~Viola are with King Abdullah University of Science and Technology (KAUST), Saudi Arabia. E-mail: \{dawar.khan, xinyu.liu,  omar.mena, donggang.jia, alexandre.kouyoumdjian, ivan.viola\}@kaust.edu.sa.}

\thanks{D. Khan and X. Liu are joint first authors with equal contributions.}
\thanks{Corresponding author: Dawar Khan}
}


\markboth{Accepted to IEEE Transactions on Visualization and Computer Graphics (TVCG), 2026.}
{Shell \MakeLowercase{\textit{et al.}}: A Sample Article Using IEEEtran.cls for IEEE Journals}

\IEEEpubid{0000--0000/00\$00.00~\copyright~2026 IEEE}

\maketitle

\begin{abstract}
The deployment of large language models (LLMs) on extended reality (XR) devices has great potential to advance the field of human-AI interaction. In case of direct, on-device model inference, selecting the appropriate model and device for specific tasks remains challenging. 
In this paper, we present \aivaluatexr, a comprehensive evaluation framework for benchmarking LLMs running on XR devices. To demonstrate the framework, we deploy 17 selected LLMs across four XR platforms—Magic Leap 2, Meta Quest 3, Vivo X100s Pro, and Apple Vision Pro—and conduct an extensive evaluation. Our experimental setup measures four key metrics: performance consistency, processing speed, memory usage, and battery consumption. 
For each of the 68 model-device pairs, we assess performance under varying string lengths, batch sizes, and thread counts, analyzing the tradeoffs for real-time XR applications.  We finally propose a unified evaluation method based on the 3D Pareto Optimality theory to select the optimal device-model pairs from the quality and speed objectives. Additionally, we compare the efficiency of on-device LLMs with client-server and cloud-based setups, and evaluate their accuracy on two interactive tasks.
We believe our findings offer valuable insights to guide future optimization efforts for LLM deployment on XR devices. Our evaluation method can be followed as standard groundwork for further research and development in this emerging field. The source code and supplementary materials are available at~\href{www.nanovis.org/AIvaluateXR.html}{\texttt{nanovis.org/AIvaluateXR.html}}.
\end{abstract}


\section{Introduction} 

\input{content/01-introduction}
\input{content/02-Relatedwork}

\input{content/03-Method}

\input{content/04a-ExpSetup}
\input{content/04b-Results}

\input{content/05-Eval-Analysis}
\input{content/06-Discussion} 
\section*{Acknowledgments}
This research has been funded by KAUST Competitive Research Grants ORFS-CRG12-2024-6422. We thank Prof. Kiyoshi Kiyokawa from NAIST, Japan, for valuable discussions. We also thank Deng Luo and Da Li from our team at KAUST for their support and valuable input across various modules.
 \bibliographystyle{IEEEtran}  

\bibliography{Llamaxr}

\vspace{-33pt} 
\input{biography.tex}

\end{document}

%% file: commands.tex
\newcommand{\etal}{\textit{et~al.}}

\newcommand{\ie}{i.e., }
\newcommand{\wrt}{w.r.t. }

\newcommand{\llama}{\emph{Llama}\xspace}

\newcommand{\aivaluatexr}{\emph{AIvaluateXR}\xspace}

\definecolor{darkred}{RGB}{192, 0, 0}

\definecolor{darkblue}{RGB}{0, 0, 192}

%% file: acronyms.tex
\newacronym{llm}{LLM}{Large Language Model} 
\newacronym{asr}{ASR}{Automatic Speech Recognition}
\newacronym{adb}{ADB}{Android Debug Bridge}
\newacronym{gguf}{GGUF}{GPT-Generated Unified Format}
\newacronym{ui}{UI}{User Interface}
\newacronym{nn}{NN}{Neural Network}
\newacronym{pp}{PP}{Prompt Processing}
\newacronym{tg}{TG}{Token Generation}
\newacronym{bt}{BT}{Batch Test}
\newacronym{tt}{TT}{Thread Test}
\newacronym{rss}{RSS}{Resident Set Size} 
\newacronym{avp}{AVP}{Apple Vision Pro} 
\newacronym{mq3}{MQ3}{Meta Quest 3} 
\newacronym{ml2}{ML2}{Magic Leap 2} 
\newacronym{vivo}{Vivo}{Vivo X100 Pro}

%% file: content/01-Introduction.tex
\IEEEPARstart{E}{xtended Reality (XR)} and Artificial Intelligence (AI) have become increasingly prominent, and their intersection is even more compelling for researchers and real-world applications~\cite{HirzleXRandAIReview,Zheng2025Reviewcompusurvey,Tang2025CHIReview,DEVAGIRI2022ARAIReview,Bozkir2024Review:LLMXR,Suzuki2025chiWorkshop}.
Since the 2022 release of OpenAI's ChatGPT interface \cite{ChatGPT-Hystor_2023}, relying on the GPT-3.5 \gls{llm}, we have witnessed the disruptive and transformative effects of \glspl{llm}.  
These models are capable of describing a wide variety of topics, respond at various levels of abstraction, and communicate effectively in multiple languages. 
They have proven capable of providing users with accurate and contextually appropriate responses. 
In addition to general applications such as grammar and text processing~\cite{Eker-GrammarCorrection_2023,Adeshola-ChatGPT-Edu_2023}, automated chatbot services, and source code generation~\cite{Feng-Software-ChatGTP_2023}, \glspl{llm} are increasingly being utilized in computer vision, human-computer interaction, and visual understanding.

In computer vision, language models are combined with visual signals to perform tasks such as verbal scene description and open-world scene graph generation~\cite{Koch2024}.  
In user interaction and visualization research, \glspl{llm} serve as verbal interfaces for controlling software functionality or adjusting visualization parameters~\cite{Jia2024_1,Jia2024_2}.  
Another disruptive application of \glspl{llm} involves their use in wearable technology~\cite{Srinidhi2024ISMAR:XaiR} to assist users in various environments. For instance, Meta AI is being integrated into Meta's virtual reality hardware, where images captured by the device are streamed over the network to Meta's servers for model inference, enabling image interpretation or prompt response tasks\footnote{https://www.meta.com/blog/quest/meta-ai-on-meta-quest-3/ \\}. The overarching vision is to develop assistive hardware that is as lightweight as sunglasses but capable of aiding users in a wide range of tasks by understanding the scene they are viewing.

 While such technological advances have the potential to empower users in unprecedented ways, the need for on-device AI in XR arises in several scenarios: (i) regions with limited or no internet connectivity (e.g., field training or remote maintenance); (ii) strict security and privacy requirements (e.g., healthcare, defense); (iii) reducing network latency; (iv) enabling autonomous operation as both LLMs and XR hardware continue to advance rapidly; (v) avoiding subscription costs; and (vi) providing the ability to fine-tune models for specific tasks and domains.
In this regard, several smaller \glspl{llm} have been released by Meta, Mistral AI, Microsoft, and Google~\cite{Liu2024}. These models are designed to operate efficiently on mobile devices and extended reality hardware, typically consisting of billions of trainable parameters, while some \glspl{llm} with fewer than a billion parameters have already demonstrated promising performance~\cite{Liu2024}.

Local execution of \glspl{llm} on XR devices is expected to become increasingly necessary for a wide range of applications. However, selecting the optimal device and model for a specific application is a complex decision. Device specifications and model documentation alone are not sufficient for making this decision. Furthermore, the hardware configurations of XR devices (such as CPU, memory capacity, and thermal management) and the architectures of the LLMs can significantly influence the results. Therefore, defining standardized evaluation criteria that address these complexities and ensure fairness across tests is also a research challenge. To the best of our knowledge, no comprehensive study currently provides clear guidelines for such evaluations. 

To address these challenges, we present \aivaluatexr, a comprehensive framework for deploying LLMs on XR platforms and systematically evaluating their performance across multiple metrics. \aivaluatexr works in two steps: first, it deploys LLMs on XR devices and evaluates their performance across a range of factors; then, in the second step, it applies Pareto analysis to the results to identify the best-performing model-device pairs.
We applied \aivaluatexr to deploy 17 \glspl{llm} across four XR devices—Magic Leap 2, Meta Quest 3, Vivo X100s Pro\footnote{Vivo X100s Pro is, strictly speaking, not an XR device. We include it in our comparison because smartphones are commonly used as XR platforms, and this device features competitive hardware specifications.}, and Apple Vision Pro—and conducted a thorough evaluation in terms of performance consistency (i.e., stability over time), processing speed, memory usage, and battery consumption. Finally, we used Pareto analysis to identify the best-performing model-device pairs among the 68 combinations evaluated.\\
We selected these four devices because they span the current spectrum of commercially available XR-capable hardware, differing in CPU architecture, memory capacity, thermal design, and support for on-device inference.  
Similarly, the seventeen evaluated LLMs (see~\Cref{tab:models}) were chosen to cover multiple model families (Qwen, Vikhr-Gemma, Phi-3.1, LLaMA-2, and Mistral-7B), parameter scales ranging from 0.5B to 7.5B, and commonly used quantization levels, ensuring a representative and practical benchmark for on-device XR deployment.

Considering the multiple challenges and factors affecting the evaluation, we first identify key evaluation metrics, and then design an experimental setup to minimize potential bias.  A key challenge is the effect of varying string lengths, batch sizes, thread counts, and background processes. To account for string length variation, we conducted experiments using different lengths. To mitigate hidden factors like background processes, each experiment was repeated five times, and average values were calculated. To provide task-specific analysis, experiments were conducted across all \glspl{llm} tasks: \gls{pp} for evaluating performance with varying prompt lengths, \gls{tg} for varying generated token set sizes, plus \gls{bt} and \gls{tt} for assessing parallelism and concurrency efficiency.
We believe our experimental protocol and results will serve as a foundation for further research and development in this emerging field. In summary, this paper contributes: 
\begin{itemize}[noitemsep,leftmargin=8pt, topsep=0pt]
\item An evaluation framework and an experimental setup for assessing the performance of \glspl{llm} on XR devices, which can be used as standard  guidelines for future research.
\item The deployment of various \glspl{llm} on XR devices, along with an analysis of their accuracy for interactive applications using standard query datasets, and a comparison with cloud-based and client-server-based LLMs in terms of efficiency and accuracy.
\item A comprehensive benchmark of 17 LLMs on four XR devices (68 model–device pairs), revealing key insights:  
(1) smaller models consistently achieve higher processing speeds and better energy efficiency;  
(2) device performance differs substantially due to CPU architecture, thermal behavior, and memory limits, with AVP generally achieving the highest processing speed, followed by ML2, Vivo, and MQ3 (though MQ3 outperforms Vivo in TG);  
(3) repeated runs show noticeable variability, indicating stability challenges for on-device inference;  
(4) model size strongly correlates with performance: larger models ($m_{12}$-$m_{17}$, LLaMA-2 and Mistral-7B) are the slowest, while smaller models such as m1, m2, m3, and m5 achieve the best throughput.
\item A cross-device analysis of speed, memory usage, energy consumption, and accuracy, with comparisons to client-server and cloud-based pipelines, highlighting practical guidelines for selecting suitable model–device combinations for real-world XR applications.
\end{itemize}
The remainder of the paper is structured as follows. 
Section~\ref{sec:related-work} reviews prior work on on-device and XR-based LLM evaluation. 
Section~\ref{sec:method} introduces the overall methodology, including the deployment pipeline, initial investigation, and evaluation metrics. 
Section~\ref{sec:exp:design} details the experimental setup and measurement procedures. 
Section~\ref{sec:results} presents the benchmarking results across all metrics. 
Section~\ref{sec:analysis} provides additional analyses of on-device \glspl{llm}, including evaluations on interactive XR datasets and a latency comparison across deployment modes (on-device vs.\ cloud-based vs.\ client–server). 
Section~\ref{sec:discussion} discusses implications, design considerations, and limitations. 
Finally, Section~\ref{sec:conclusion} concludes the paper and outlines directions for future work.

%% file: content/02-Relatedwork.tex
\section{Related Work}
\label{sec:related-work}  
We review related work along two major research directions: LLM-powered XR and on-device LLMs. For further details on AI in XR, see the recent review articles~\cite{HirzleXRandAIReview,Zheng2025Reviewcompusurvey,DEVAGIRI2022ARAIReview,Tang2025CHIReview}.
\subsection{LLM-powered XR Applications}
Owing to the powerful semantic understanding and extensive general knowledge of LLMs, numerous studies have explored their application in assisting various tasks within XR scenarios. Recent studies have demonstrated the use of LLMs for natural interaction and conversational interfaces in VR and AR scenarios~\cite{Bellucci2025VRLLMs,Chen2025vrLLMs,Afzal2025NextXRLLMCGA}. LLMs have also been applied for content generation, chat-based assistance, VR education\cite{Gao2025PerVRML}, social VR~\cite{wan2024CHISocailVR}, and as intelligent companions in XR applications~\cite{Li2025LLM4VR,Li2025CHI}.
In addition to LLMs, vision-language models (VLMs) have been integrated into AR systems to enhance visual and multimodal understanding~\cite{Srinidhi2024ISMAR:XaiR}, scene-driven AR storytelling~\cite{Li2022ARstoryTOG,sun2025vlmStory}, contextual guidance, dynamic content creation, and much more~\cite{Rubinstein2024ObjectAIxVR}.

\textit{XaiR}~\cite{Srinidhi2024ISMAR:XaiR} integrates Multimodal Large Language Models (MLLMs) with XR devices using a client-server architecture, where computationally intensive MLLM tasks are offloaded to a server while spatial context is handled locally on the XR headset. The system supports real-time multimodal input; however, latency comprising both network delays and LLM inference time remains a challenge, often requiring users to wait 4--6 seconds for a response. Torre \etal\cite{de2024llmr} introduced a similar framework, \emph{LLMR}, which leverages LLMs for real-time creation and modification of interactive mixed reality experiences, enabling tasks such as generating new content or editing existing elements on VR/AR devices.

Jia \etal~\cite{Jia2024_1} developed the \emph{VOICE} framework, which employs a two-layer agent system for conversational interaction and explanation in scientific communication, with a prototype deployable on VR devices. Kurai \etal~\cite{kurai2024magicitem} proposed \emph{MagicItem}, a tool that allows users with limited programming experience to define object behaviors within the metaverse platform. Zhang \etal~\cite{zhang2024odoragent} introduced \emph{OdorAgent}, which combines an LLM with a text-image model to automate video-odor matching. Yin \etal~\cite{yin2024text2vrscene} identified potential limitations in LLM-based automated systems and proposed the systematic framework \emph{Text2VRScene} to address them. Chen \etal~\cite{chen2024supporting} leveraged the extensive capabilities of LLMs in context perception and text prediction to enhance text entry efficiency by reducing manual keystrokes in VR scenarios. Giunchi \etal~\cite{giunchi2024dreamcodevr} developed \emph{DreamCodeVR}, a tool designed to help users, regardless of coding experience, create basic object behavior in VR environments by translating spoken language into code within an active application. Wan \etal~\cite{wan2024building} presented an LLM-based AI agent for human-agent interaction in VR, involving GPT-4 to simulate realistic NPC behavior, including context-aware responses, facial expressions, and body gestures.

XR applications often require 3D content that can be generated using AI. For example, Kawka \etal~\cite{Kawka2025web} proposed a method to create 3D objects from textual and image data, along with techniques for surface remeshing and simplification. Other approaches include character animation~\cite{Clocchiatti2024CharacterAnimationAIxVR}, text-to-3D~\cite{Ming2024Instant3D}, speech-to-3D generations~\cite{Weng2025DreamMeshAIxVR}, and image-to-3D reconstruction~\cite{Liu2023zero123}. Recent work has also explored modeling human avatars and motion~\cite{Chen2025MDDAIxVR}, as well as spatial understanding and object recommendation~\cite{Behravan2025AIxVRVLM43D}.
Although these systems typically rely on cloud-based models or client-server architectures rather than performing inference directly on the device, they demonstrate a clear need for AI in XR. This dependence raises concerns about user privacy, latency, costs, and internet access requirements. In our work, we evaluate the local inference performance of various LLMs and propose a prototype capable of running LLM inference entirely on-device.

\subsection{On-device LLMs}
Running LLMs on edge devices, commonly called on-device LLMs, has garnered significant research interest due to their advantages in enhancing privacy, reducing latency, and operating without the need for internet connectivity. Because of the limited memory and computing capabilities, on-device LLMs usually require resource-efficient
LLM deployment~\cite{qu2024mobile}: a trade-off between performance and model size.

Cheng \etal~\cite{cheng2023optimize} introduced the \emph{SignRound} method, which leverages signed gradient descent to optimize both rounding values and weight clipping. This approach achieves outstanding performance in 2- to 4-bit quantization while maintaining low tuning costs and eliminating additional inference overhead. Ma \etal~\cite{ma2023llm} developed the \emph{LLM-Pruner} method, which uses structural pruning to selectively remove non-essential coupled structures based on gradient information, effectively preserving the core functionality of the LLM. Their results demonstrate that the compressed models continue to perform well in tasks such as zero-shot classification and generation. Gu \etal~\cite{gu2024minillm} introduced a novel knowledge distillation method that compresses LLMs into smaller models, resulting in the student model \emph{MINILLM}. \emph{MINILLM} demonstrates superior performance compared to baseline models, producing more precise responses with reduced exposure bias, improved calibration, and enhanced long-text generation capabilities. Liu \etal~\cite{liu2024kivi} developed a tuning-free 2-bit KV cache quantization algorithm, named \emph{KIVI}, which independently quantizes the key cache per channel and the value cache per token. This approach enabled up to a 4× increase in batch size, resulting in a $2.35\times$ to $3.47\times$ improvement in throughput on real LLM inference workloads. Liu \etal~\cite{liu2024mobilellm} introduced \emph{MobileLLM}, which explores the importance of model architecture for sub-billion-parameter LLMs by leveraging deep and narrow architectures, combined with embedding sharing and grouped-query attention mechanisms. 

Although these studies primarily focused on reducing model size efficiently for deployment on various devices, they did not evaluate the models' performance in real-world on-device scenarios, which is the focus of our paper.

%% file: content/03-Method.tex
\section{Research Methods and Materials} 
\label{sec:method}    
This section outlines our research methodology, including the key challenges, the LLM deployment approach, and the evaluation metrics and models used in this study.  

 \textbf{Problem Statement:}
Given a set of \glspl{llm} $\mathcal{M} = \{m_i\}_{i=1}^N$ and a set of XR devices $\mathcal{D} = \{d_j\}_{j=1}^K$, \aivaluatexr evaluates the performance of each model-device pair \((m_i, d_j)\) across several key factors and quantitative metrics. It first generates evaluation results for each metric, which are then passed to a Pareto analysis module to determine the Pareto-optimal set.
We used \aivaluatexr to deploy 17 \glspl{llm}, denoted as \( \{ m_1, m_2, \dots, m_{17} \} \) (see~\cref{tab:models}), across four XR and mobile devices: \( d_1, d_2, d_3, \) and \( d_4 \), representing \gls{ml2},\footnote{\url{https://www.magicleap.com/magicleap-2}} \gls{mq3},\footnote{\url{https://www.meta.com/quest}} \gls{vivo},\footnote{\url{https://www.vivo.com/en}} and \gls{avp},\footnote{\url{https://www.apple.com/vision-pro}} respectively. Our goal is to perform a comprehensive evaluation of these models on the four devices to determine how well they handle large-scale language processing tasks in resource-constrained XR environments. 

\begin{figure}[tbp]
    \centering   
    \includegraphics[width=\linewidth]{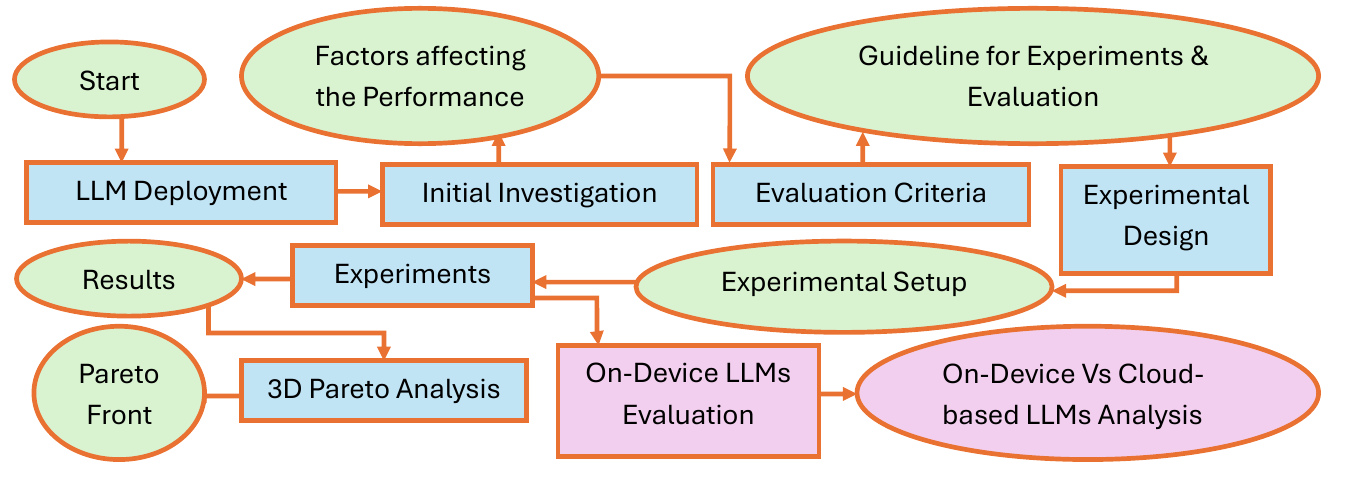}  
\vskip -0.245cm \caption{Overview of our research process. The rectangular boxes represent the procedures, while the ovals indicate the outcomes. We deployed 17 \glspl{llm} on four XR devices. First, we identified the key factors affecting performance. Based on these factors, we defined our evaluation criteria and conducted the experiments. Finally, we applied 3D Pareto analysis, which provided the Pareto front to highlight the best-performing model-device pairs. We also analyzed on-device LLMs versus cloud-based LLMs (highlighted in pink).}
    \label{fig:mainMethod}
    \vskip -0.5cm 
\end{figure}

We define our evaluation using six key performance metrics: 
\( p_0, p_1, p_2, p_3, p_4, p_5 \), which represent  
\textit{Model Quality},  
\textit{Performance Consistency},  
\textit{Processing Speed} (\wrt string length and prompt length),  
\textit{Parallelism} (\wrt thread count and varying batch size),  
\textit{Memory Usage}, and  
\textit{Battery Usage}, respectively.
\cref{fig:mainMethod}  presents an overview of our research pipeline, where each rectangular box represents a specific procedure, and the connected ovals indicate their respective outcomes. These are discussed in detail below.
\subsection{LLM Deployment on XR Devices}
We aim to deploy the \glspl{llm} locally on XR devices. For this purpose, we customize the Llama.cpp library \cite{meta2023introducingLLaMA} and build it for the four target XR devices. The resulting application is capable of loading various appropriately sized \gls{gguf} models.  Our deployment leverages the functionalities of \llama, allowing us to run basic \glspl{llm} tasks, and adjust various parameters.
For example, to evaluate different aspects in an experiment, we can control various parameters such as prompt length in \emph{Prompt Processing} (\gls{pp}) and the size of the token set in \emph{Token Generation} (\gls{tg}). To benchmark the models on various devices, the top-level scripts in Llama.cpp are compiled into binary executable files. These binaries include scripts specifically designed for testing both execution speed and model quality. For Apple Vision Pro, we utilized Xcode\footnote{\url{https://developer.apple.com/xcode/}}, Apple's official IDE, to build and deploy testing scripts directly to the device. For the remaining three devices, the testing was performed via a shell interface opened through \gls{adb}.
In this study, all 17 selected models were deployed on each of the four devices.

Due to compatibility constraints with Llama.cpp, we conducted testing exclusively on the CPU for Magic Leap 2, Meta Quest 3, and Vivo X100s Pro. As a smartphone, the Vivo X100s Pro theoretically supports GPU inference through frameworks like TensorFlow Lite or ONNX Runtime, but Llama.cpp does not currently offer GPU inference support for this device. Similarly, Magic Leap 2 and Meta Quest 3, being XR-specific devices, lack user-accessible GPU inference capabilities compatible with Llama.cpp. In contrast, Apple Vision Pro supports GPU inference via Metal, which we used in our experiments. These device and framework limitations necessitated restricting inference to CPUs for the first three devices.
\subsection{Initial Investigation and Factors Identification} 
\label{sec:InitInvestigation}
 After deploying the \glspl{llm}, 
 we conducted an initial investigation through anecdotal informal tests, literature reviews, and by studying LLM documentations. The aim of this preliminary investigation was to identify the key challenges and factors affecting performance---prerequisites for a fair evaluation---and to define our evaluation approach (\cref{sec:evCriteria}).
 
 We observed that XR devices experience performance degradation during prolonged tasks due to factors such as overheating, battery depletion, and background processes. Moreover, they often do not perform at a stable level. For instance, with the same model-device pair and same parameters (input settings), we obtained different results across different runs of the same test. We refer to the metric used for measuring this variability as \textit{Performance Consistency}. 
 This variability mainly stems from system-level factors such as 
background OS activity, thermal throttling, power-management 
adjustments, and initialization overheads during LLM inference. These 
factors cause slight fluctuations even when the model, device, and 
parameters remain identical across runs.

 Another important factor is execution time: the time taken to execute a particular task. In our context, we used a relative term called \textit{Processing Speed}, measured in \textit{tokens per second}. We observed variations in processing speed depending on varying  string length, batch size, and thread count. In addition, \textit{Memory Consumption} and \textit{Battery Consumption} are also important factors to consider. Similarly, we observed different processing speeds for different tasks. For example, prompt processing and token generation exhibited varying performance, even with the same model-device pairs and parameters.

In addition to the parameters mentioned above, during the initial investigation, we also noticed that ensuring consistent testing conditions across devices is critical for fair comparisons but challenging to maintain. For example, variance in a device's \textit{Performance Consistency} could create biases into the results. LLM performance may vary due to different architectures and device-specific optimizations. Even with identical parameters, the same device and LLM can produce different results due to unknown factors. The device's environment, such as exposure to heat or obstructions to the cooling fan, can also affect performance. The concise presentation of the findings, and ensuring their reproducibility for re-evaluation and validation, is an additional challenge. Considering these challenges, we formulated a set of guidelines as our evaluation approach in~\cref{sec:evCriteria}. Based on the outcome of this investigation and to ensure full transparency of the evaluated models, Table~\ref{tab:models} summarizes all 17 LLMs used in our study.

\begin{table*}[!htb]
\caption{Models used, with their IDs, names, versions (v), quantization level (Q), layers (L), parameters (P), and sizes, sourced from Hugging Face~\cite{huggingface2024}.}
\vskip -0.1345cm 
\centering
\small
\begin{tabular}{l l l c c c c c}
\hline
Series & ID & Model Name  & v & Q & L & P & Size \\
\hline
\multirow{1}{*}{Qwen} & $m_1$   & qwen2-0\_5b-instruct-fp16  & v2.0 & FP16 & 24 & 0.5B & 0.942 GB \\ 

 \hline
\multirow{4}{*}{Vikhr-Gemma}
& $m_2$  & Vikhr-Gemma-2B-instruct-Q3\_K\_M & v1.3 & Q3 & 26 & 2.61B & 1.36 GB \\ 
& $m_3$ & Vikhr-Gemma-2B-instruct-Q4\_0 & v1.3 & Q4 & 24 & 2.61B & 1.51 GB \\ 
& $m_4$ & Vikhr-Gemma-2B-instruct-Q5\_0  & v1.3 & Q5 & 24 & 2.61B & 1.75 GB \\ 
& $m_5$  & Vikhr-Gemma-2B-instruct-Q6\_K  & v1.3 & Q6 & 24 & 2.61B & 2.00 GB \\ 

\hline
\multirow{6}{*}{Phi-3.1}  
& $m_6$ & Phi-3.1-mini-4k-instruct-Q2\_K  & v3.1 & Q2 & 20 & 3.82B & 1.32 GB \\ 
& $m_7$  & Phi-3.1-mini-4k-instruct-Q3\_K\_L  & v3.1 & Q3 & 20 & 3.82B & 1.94 GB \\ 
& $m_8$  & Phi-3.1-mini-4k-instruct-Q4\_K\_L &  v3.1 & Q4 & 20 & 3.82B & 2.30 GB \\ 
& $m_9$ & Phi-3.1-mini-4k-instruct-Q5\_K\_L & v3.1 & Q5 & 20 & 3.82B & 2.68 GB \\ 
& $m_{10}$ & Phi-3.1-mini-4k-instruct-Q6\_K & v3.1 & Q6 & 20 & 3.82B & 2.92 GB \\ 
& $m_{11}$  & Phi-3.1-mini-4k-instruct-Q8\_0  & v3.1 & Q8 & 20 & 3.82B & 3.78 GB \\ 

\hline
\multirow{2}{*}{LLaMA-2}  
& $m_{12}$  & llama-2-7b-chat.Q2\_K &  v2.0 & Q2 & 28 & 6.74B & 2.63 GB \\ 
& $m_{13}$  & llama-2-7b-chat.Q3\_K\_S  & v2.0 & Q3 & 28 & 6.74B & 2.75 GB \\ 

\hline
\multirow{4}{*}{Mistral-7B}  
& $m_{14}$ & Mistral-7B-Instruct-v0.3.IQ1\_M  & v0.3 & IQ1 & 28 & 7.25B & 1.64 GB \\ 
& $m_{15}$  & Mistral-7B-Instruct-v0.3.IQ2\_XS  & v0.3 & IQ2 & 28 & 7.25B & 2.05 GB \\ 
& $m_{16}$  & Mistral-7B-Instruct-v0.3.IQ3\_XS  & v0.3 & IQ3 & 28 & 7.25B & 2.81 GB \\ 
& $m_{17}$  & Mistral-7B-Instruct-v0.3.IQ4\_XS  & v0.3 & IQ4 & 28 & 7.25B & 3.64 GB \\ 

\hline
\end{tabular}
\label{tab:models} 
\end{table*}

\subsection{Evaluation Approach}
\label{sec:evCriteria}
Considering the key challenges identified in our initial investigation (see~\cref{sec:InitInvestigation}), we adopt a standardized evaluation approach that defines how \aivaluatexr assesses model--device performance in a fair, comparable, and reproducible manner. Our approach is guided by three principles:
\begin{enumerate}[noitemsep,leftmargin=10pt]
    \item \textbf{Consistency across devices:}
    All evaluations follow a unified set of conditions so that hardware or environmental differences do not introduce bias into the results.
    \item \textbf{Multi-metric assessment:}
    Since on-device LLM behavior is multidimensional, the framework evaluates several complementary metrics, including model quality, performance stability, processing speed, memory usage, and energy consumption.
    \item \textbf{Unified interpretation of results:}
    Because these metrics operate on different scales, we later integrate them through Pareto analysis to identify model--device pairs that offer optimal trade-offs.
\end{enumerate}
The specific experimental configuration including models, devices, test parameters, and measurement procedures are described in~\cref{sec:exp:design}.
\subsection{Evaluation Metrics}
\label{sec:metricsUsed} 
Our evaluation comprises multiple tests designed to assess a wide range of performance characteristics. To ensure uniform conditions, we use synthetic prompts automatically generated by \texttt{llama.cpp}~\cite{meta2023introducingLLaMA}.  The string length is controlled via a parameter set to 64, 128, 256, 512, and 1024 tokens.
The evaluation includes the following tests. 
\begin{itemize}[nolistsep]
    \item \textbf{Performance Consistency:} Measures stability over time.
    \item \textbf{Processing Speed:} Analyzes performance across different string and prompt lengths, further divided into:
    \begin{itemize}[nolistsep]
        \item \textbf{PP (Prompt Processing):} Evaluates encoding efficiency in handling input prompts of varying lengths.
        \item \textbf{TG (Token Generation):} Evaluates the device's speed in generating output tokens of varying lengths.
    \end{itemize}
    \item \textbf{Parallelization and Concurrency:} This includes:
    \begin{itemize}[noitemsep, leftmargin=8pt, topsep=0pt]  
    \item \textbf{Batch Test (BT)}: Evaluates the device's ability to process multiple input samples simultaneously by handling different batch sizes during token generation.  
    \item \textbf{Thread Test (TT)}: Measures performance scalability by varying thread counts during LLM execution.  
\end{itemize}  
\end{itemize} 
Here, \textbf{Batch Size} refers to the number of input samples processed in parallel. While larger batch sizes improve computational efficiency, they also demand more memory. Each of these metrics requires task-specific analysis for both models and devices. Furthermore, we conducted additional evaluations for \textit{Memory Usage} and \textit{Battery Consumption}. Finally, we performed a \textbf{Pareto Optimality} analysis to collectively examine the results across different models and devices.

%% file: content/04a-ExpSetup.tex
\section{Experimental Setup}
\label{sec:exp:design}
This section describes the experimental setup and methodology for evaluating processing speed and error rates. The results are presented in \cref{sec:results}. Before diving into the detailed setup, we first discuss the two evaluation metrics: processing speed and error count.

\noindent \textbf{Processing Speed:}
In this study, we measure the processing speed for each model-device pair \((m_i, d_j)\) by recording the time \(t_{ijk}\) taken by model \(m_i\) running on device \(d_j\) to process a string of length \(L\). The instantaneous processing speed, \(S_{inst.}\), expressed in tokens per second, is calculated as:
$S_{inst.} = \frac{L}{t_{ij}}$. 
However, instead of relying on a single run, we perform 5 runs for each model-device pair to ensure accuracy and measure stability. If the coefficient of variance (CV) across the 5 runs is less than \(33\%\), we calculate the 5-Run-Mean Speed (referred to as consistent speed, \(S_c\)) as:

\begin{equation}
\label{eq:mean_speed}
S_c = \frac{1}{5} \sum_{k=1}^{5} S_{inst.}(k),
\end{equation}
where \( S_{\text{inst.}}(k) \) represents the instantaneous speed during the \( k^{\text{th}} \) run of each experiment.  This approach minimizes the impact of factors such as initialization delays, hardware fluctuations, or other environmental inconsistencies, providing a more reliable measure of processing speed.\\
\noindent \textbf{Error Count:}
If the CV exceeds \(33\%\), the run is considered unstable, and the corresponding data point is excluded from further analysis. The count of such unstable runs is recorded as an error during the experiment.\\
We conducted the following experiments.
\begin{figure}[tb]
    \centering 
    \includegraphics[width=1.0\linewidth]{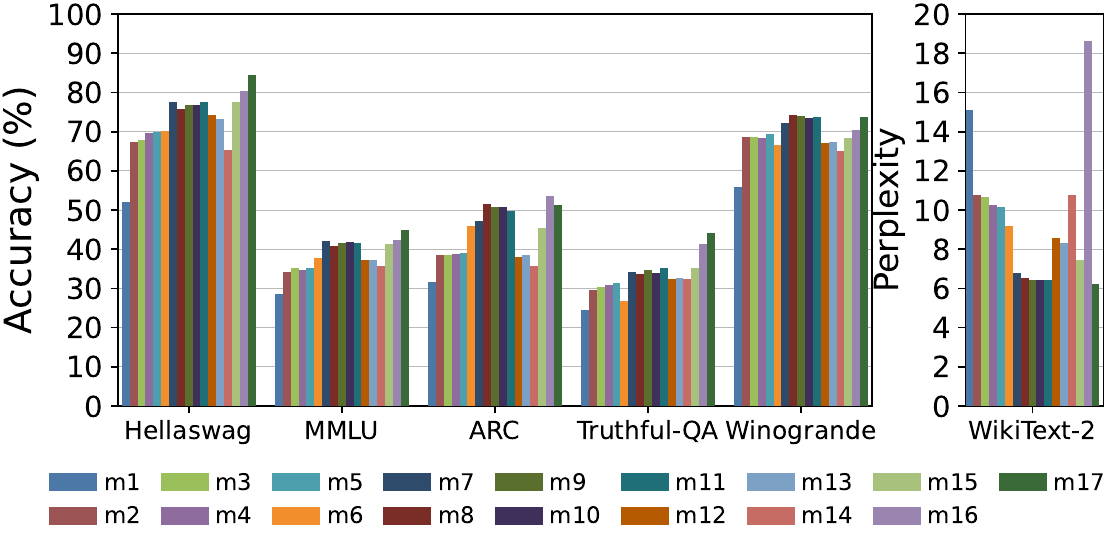}
    \vskip -0.3478cm
 \caption{Model Quality Analysis: The models were evaluated based on six different benchmarks. We chose accuracy (\textbf{$\uparrow$}) for the five benchmarks on the left and perplexity (\textbf{$\downarrow$}) for the WikiText-2 benchmark.  Each group in the bar chart represents models $m_1$ to $m_{17}$ from left to right.}
    \label{fig:Modelsqty}
\end{figure} 
\subsection{Models Quality Analysis}
\label{sec:modelsQA}
Due to the storage and computational limitations of edge devices, deploying \gls{llm} locally often requires a trade-off between model size and performance. To put our performance results in perspective, we first conducted an empirical analysis of the models we have used in this study. This section offers insights into the models evaluated in this work, presenting a comparative analysis. While this section is not directly related to device performance, we believe providing detailed information about the models is essential for understanding their capabilities and limitations.
We evaluate each model using six benchmarks, assessing various reasoning and knowledge capabilities: 
\textit{Hellaswag}~\cite{hellaswag} (commonsense reasoning), 
\textit{MMLU}~\cite{hendrycks2020measuring} (multi-subject knowledge), 
\textit{ARC}~\cite{clark2018thinksolvedquestionanswering} (complex science questions), 
\textit{TruthfulQA}~\cite{lin2021truthfulqa} (truthfulness in responses), 
\textit{Winogrande}~\cite{ai2:winogrande} (pronoun resolution), and 
\textit{WikiText-2}~\cite{merity2016pointer} (language modeling and text generation).
 
For the first five benchmarks, which consist of multiple-choice questions, we use accuracy as the evaluation metric. However, for WikiText-2, the task is to predict the probability of each word in a given text, reflecting the model's natural language understanding. Therefore, we use perplexity as the evaluation metric, defined as:
\begin{equation} \text{Perplexity}(P) = e^{-\frac{1}{N} \sum_{i=1}^{N} \log P(w_i)}, \end{equation} 
where $P(w_i)$ represents the probability assigned by the model to the $i$-th word, and $N$ is the total number of words in the sequence.
\subsection{Evaluating Performance Consistency}
\label{sec:setStability}
Some devices may experience performance degradation over time. In these experiments, we evaluate how processing speed fluctuates across multiple runs. For each model \( m_i \) and device \( d_j \), we perform \( N \) runs and compute the mean $(\mu_{ij})$ of the speed $(S_c)$, standard deviation \( \sigma_{ij} \), and coefficient of variation ($CV_{ij}$). We also record the error count, as well as the maximum and minimum speeds observed during each experiment.
In our experiments, we set \( N = 20 \), \ie each model-device pair is tested 20 times to evaluate performance variance. We noted  the 5-Run-Mean speed \( S_c \) measured during each run, calculated as in~\cref{eq:mean_speed}. In other words, each \( S_c \) itself is the mean of 5 runs, ensuring that the measured value is not too affected by errors or hidden factors.

To let the devices cool down and conduct each run in comparable circumstances, we leave each device idle for two minutes between each run of the same model. We find this to be adequate, but just to be safer, between different models, we leave each device idle for ten minutes. 
For each test, we conducted both PP and TG evaluations with the string length set to 64, \ie \( \text{PP} = 64 \) and \( \text{TG} = 64 \). For each model-device pair, we calculated the standard deviation, mean, and CV. The results are presented in~\cref{sec:resultsStability}. 

To mitigate performance fluctuations, each experiment was repeated multiple times, and the average results were computed.  
Finally, we unified the five evaluation metrics, \ie $[P_1, P_5]$, using \textit{Pareto Efficiency} theory to identify Pareto-optimal model-device pairs. Our evaluation framework and experimental design aim to facilitate reproducibility for other researchers. Additionally, we will provide open-source test scripts to enable replication of our study and further evaluations.
\subsection{Processing Speed and String Length}
\label{sec:settimePPTG}
String length, comprising both \textit{Prompt Length} and \textit{Size of the Token Set}, significantly affects performance. We evaluate the performance of LLMs on XR devices with varying string lengths to assess their processing speed. For each experiment, the processing speed is calculated as in~\cref{eq:mean_speed}.

We conducted experiments for five different string lengths (64, 128, 256, 512, and 1024) under the following PP and TG setups.
\begin{itemize}[noitemsep,leftmargin=8pt,topsep=0pt]
    \item \textbf{\gls{pp}:} The PP values were set to 64, 128, 256, 512, and 1024, while TG was fixed at 0, and other parameters at their default values.
    \item \textbf{\gls{tg}:} The TG values were set to 64, 128, 256, 512, and 1024, while PP was set 0, and other parameters were at default values.
\end{itemize}
We set PP or TG to zero in these experiments to isolate the cost of each
stage independently. This ensures that the PP test results are not affected by TG, and the PP results are not affected by PP.

\begin{figure*}[!htbp]
    \centering  \includegraphics[width=0.997\linewidth]{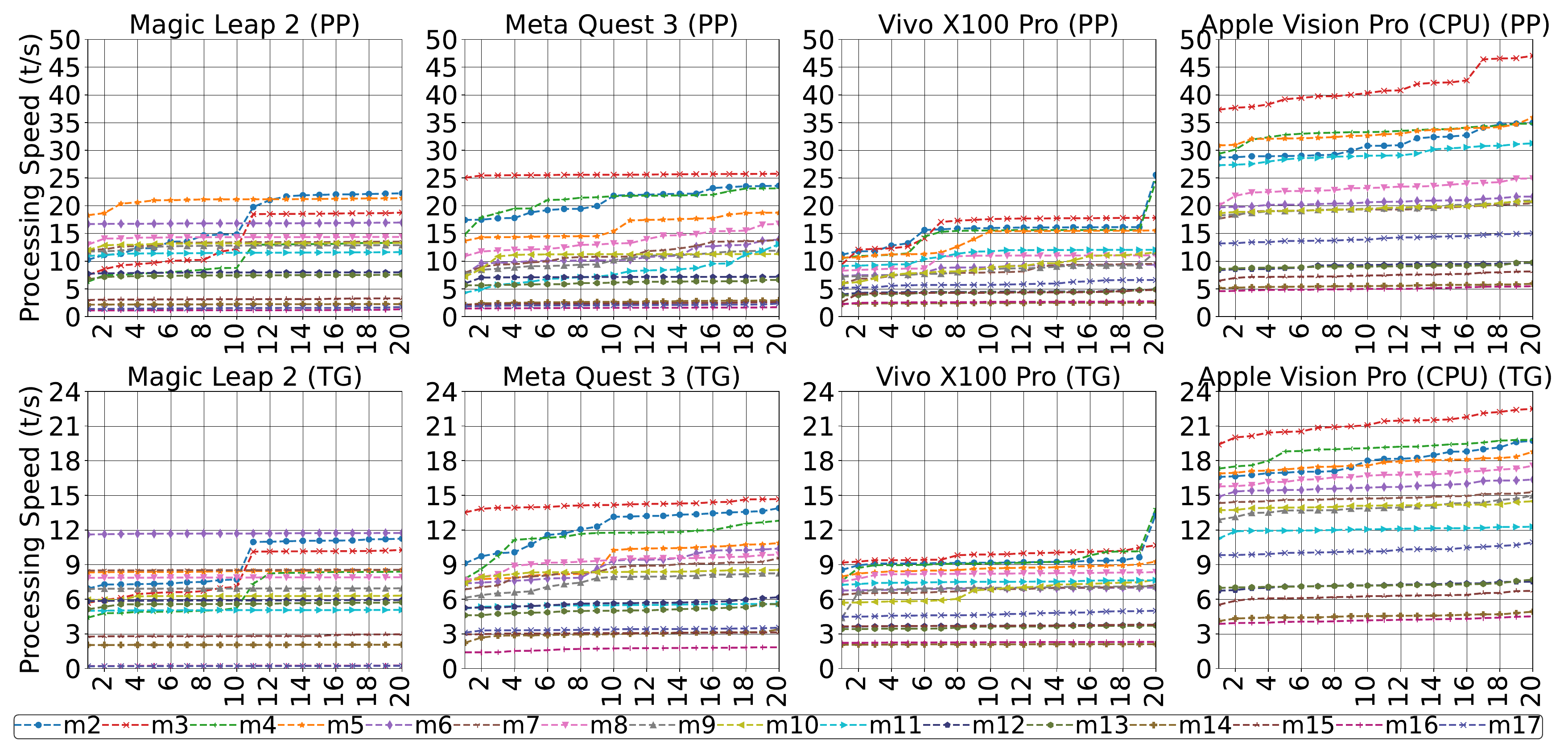}  
    \vskip -0.478cm
\caption{Consistency results of the four devices over time: PP (top) and TG (bottom) speeds in tokens per second across 20 sorted runs (X-axis: run number, Y-axis: speed in $t/s$). Model IDs follow~\Cref{tab:models}: 
Vikhr-Gemma (m$_2$–m$_5$),
Phi-3.1 (m$_6$–m$_{11}$),
LLaMA-2 (m$_{12}$–m$_{13}$),
Mistral-7B (m$_{14}$–m$_{17}$).
}
    \label{fig:StabDD}
\end{figure*}
\begin{figure}
     \includegraphics[width=1.0\linewidth]{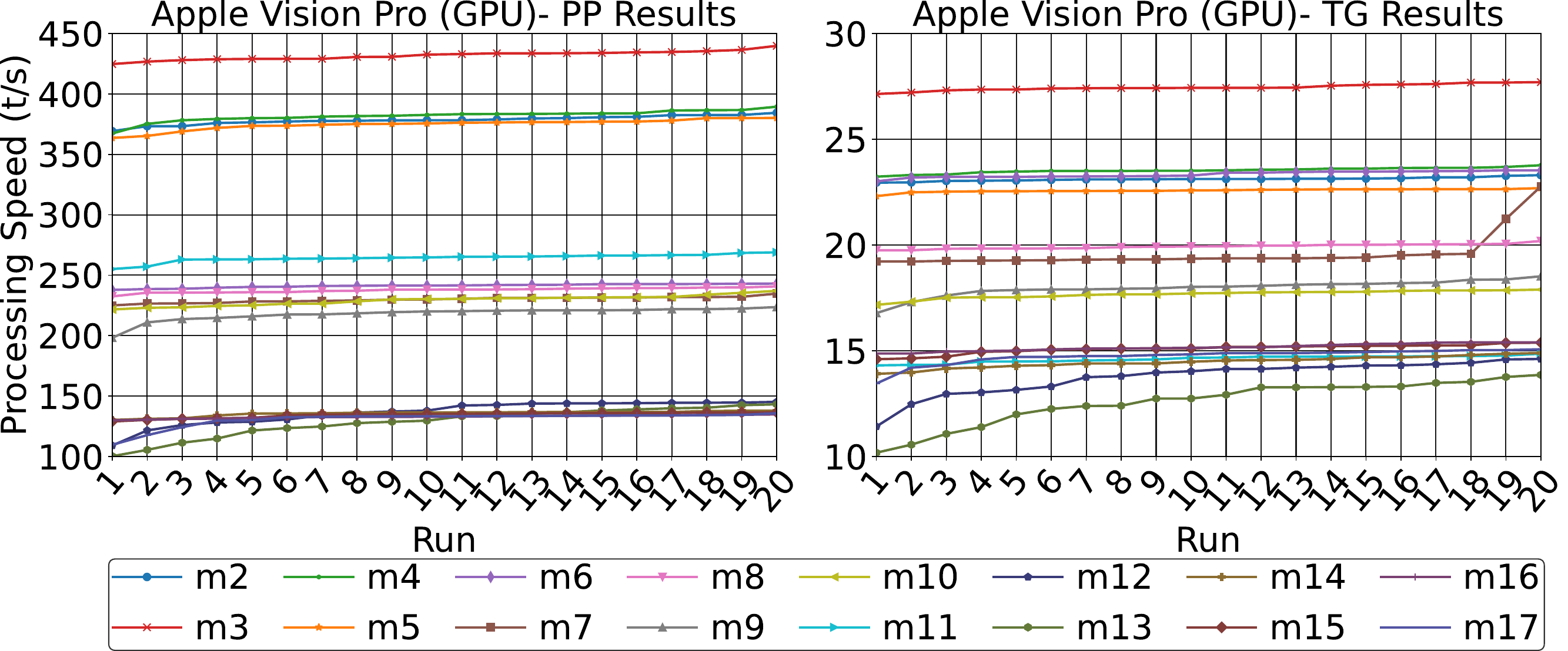}  
\vskip -0.295cm 
    \caption{Consistency results of the \gls{avp} on GPU. Left: PP results. Right: TG results. Each model was tested 20 times, and the results are plotted in sorted order. Model IDs follow~\Cref{tab:models}:
Vikhr-Gemma (m$_2$–m$_5$),
Phi-3.1 (m$_6$–m$_{11}$),
LLaMA-2 (m$_{12}$–m$_{13}$),
Mistral-7B (m$_{14}$–m$_{17}$).}
    \label{fig:VPGPU}
\end{figure}
\subsection{Parallelization with Thread Count and Batch Size}
\label{sec:setTTBT}
The parameters \textit{Batch Size} and \textit{Thread Count}, which control parallelization, also significantly affect performance. We evaluate the performance of LLMs on XR devices with varying batch sizes and thread counts to assess their processing speed, calculated using~\cref{eq:mean_speed}. The two tests conducted are described below: 
\begin{itemize}[noitemsep,leftmargin=8pt,topsep=0pt]
    \item \textbf{\gls{bt}:}  We set PP to 64, TG to 0, and varied the batch size (128, 256, 512, 1024), keeping other parameters at default values.
    \item \textbf{\gls{tt}:} We set PP to 64 and TG to 0, and varied the thread counts (1, 2, 4, 8, 16, 32), keeping other parameters at default values.
\end{itemize}
\subsection{Evaluation of Memory Consumption}
\label{sec:resultsMemory}
Memory consumption \( M_{ij} \) is measured in terms of the \gls{rss} which represents the actual physical memory usage of the relevant processes. 
For each model-device pair  ($m_i, d_j$), we measure memory usage over three runs, and report the average.
 The memory usage for PP and TG parameters was evaluated using the same approach as the performance consistency assessment. In each test, a fixed prompt was provided, and the model was instructed to generate outputs. The experiments were conducted with batch sizes of 128, 256, 512, and 1024, and their average memory consumption was recorded.   The results are shown in~\cref{sec:resultsMemory}.
\subsection{Battery Consumption}
\label{sec:resultsBattery} 
This section presents battery consumption rates during extended \gls{pp} and \gls{tg} tests. From our initial investigation, we observed that only \gls{avp} exhibited a slight impact of model size on battery consumption, while the other three devices did not show significant variations with changing model sizes. 
Given this observation, instead of testing all 17 models, we conducted battery tests on selected models. Specifically, we chose $m_1$ as the sole model from the Qwen Series. For the remaining series, we selected the smallest and largest models: $m_2$ and $m_5$ from the Vikhr-Gemma Series, $m_6$ and $m_{11}$ from the Phi-3.1 Series, $m_{12}$ and $m_{13}$ from the LLaMA-2 Series, and $m_{14}$ and $m_{17}$ from the Mistral-7B Series. 
For each experiment, we recorded the battery level at the start and end of the experiment, allowing us to calculate battery consumption over a fixed duration of $600$ seconds (10 minutes). Each experiment was repeated three times, and we reported the mean values.  To ensure consistent conditions, we took a $600$-second break between experiments, allowing the devices to cool down. The results are presented in~\cref{sec:resultsBattery}.
\subsection{Pareto Optimality}
Although we evaluate our five performance metrics across various devices and models, the metrics remain fragmented and cannot be directly compared with other metrics. To address this, we turn to \textit{Pareto efficiency} theory~\cite{chinchuluun2007survey}, which provides a measure of efficiency in multi-objective contexts and has been widely applied in various system design approaches~\cite{brisset2015approaches, santoro2018design}. A choice is considered Pareto optimal if no other choice exists that can improve any of its objective criteria without deteriorating at least one other criterion. In our case, $x_{1}$ is considered dominated by $x_{2}$ through objects $f$ if :
\begin{multline} \label{equ:pareto}
f_i(x_1) \leq f_i(x_2) \quad \forall i \in \{1, \dots, m\} \\
\text{and} \quad \exists j \in \{1, \dots, m\} \mid f_j(x_1) < f_j(x_2)
\end{multline}
where $i$, $j$ represent different objective indices.

A choice $x^*$ is considered Pareto optimal if no other feasible option dominates it. The set of all non-dominated designs forms the Pareto front, which represents the optimal trade-offs between all objectives.

To identify the optimal choices across devices and models, we define three objectives: \textit{quality}, \textit{performance}, and \textit{stability}. To calculate the final score for each objective, we propose~\cref{equ:score}. For a given device-model pair and a specific objective, we first apply min-max normalization to each metric across all device-model pairs to eliminate the impact of different scales. Then, we compute a weighted sum of all metrics for that pair to obtain a single score:
\begin{equation} \label{equ:score}
    f_{o}(x) = \sum_{i=1}^{n} w_i \cdot \frac{m_i(x) - \min(m_i)}{\max(m_i) - \min(m_i)},
\end{equation}
where $o$ is the objective index, $n$ is the number of metrics used in the objective category, $w_i$ defines the weight for each metric, and $\min$ and $\max$ calculate the minimum and maximum values across the entire set of results for metric $m_i$.
In practice, some metrics, such as perplexity, are better when lower, which is the opposite of the direction in~\cref{equ:pareto}. Therefore, we take the reciprocal of each perplexity value.  

%% file: content/04b-Results.tex
\begin{figure}
     \centering      
     \includegraphics[width=\linewidth]{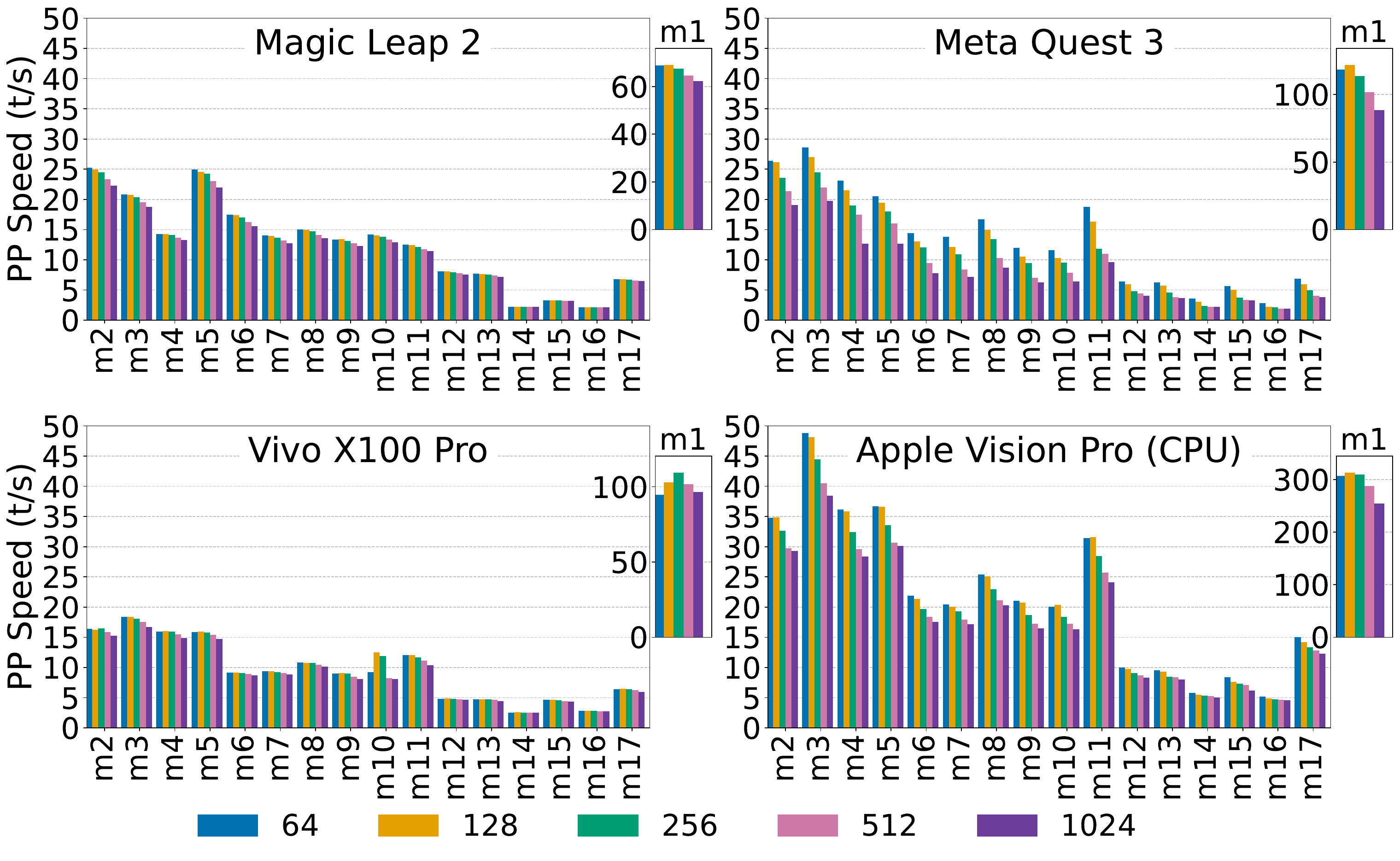}
\vskip -0.45cm 
    \caption{Processing speed of the four devices in the PP test with varying string lengths: 64, 128, 256, 512, and 1024. The x-axis represents the model, while the y-axis represents the processing speed in t/s. For $m_1$, a different scale is used due to its higher values. Model IDs follow~\Cref{tab:models}:
Qwen (m$_1$),
Vikhr-Gemma (m$_2$–m$_5$),
Phi-3.1 (m$_6$–m$_{11}$),
LLaMA-2 (m$_{12}$–m$_{13}$),
Mistral-7B (m$_{14}$–m$_{17}$).}
    \label{fig:resPP:CPU}
\end{figure}


\section{Performance Evaluation Results}
\label{sec:results}
This section presents the results of four types of experiments described in \cref{sec:exp:design}.
\subsection{Model Quality Analysis}
\label{sec:resModelsQty}
As shown in \cref{tab:models}, we select five different model architectures, and for each architecture, we choose various quantization settings. While lower-bit quantization reduces model size, speeds up inference, and lowers power consumption, it typically comes at the cost of reduced model performance. The results of the model quality are illustrated in \cref{fig:Modelsqty}.
As shown in \cref{fig:Modelsqty}, a consistent trend is observed across all benchmarks: applying lower-bit quantization settings results in reduced model performance. Models with more parameters exhibit better language understanding capabilities, leading to higher performance under the same quantization settings. Additionally, different benchmarks exhibit varying levels of sensitivity to the quantization settings. For example, Hellaswag, ARC and WikiText-2 are more sensitive compared to the other three benchmarks.

\begin{figure}[t]
     \centering      
     \includegraphics[width=\linewidth]{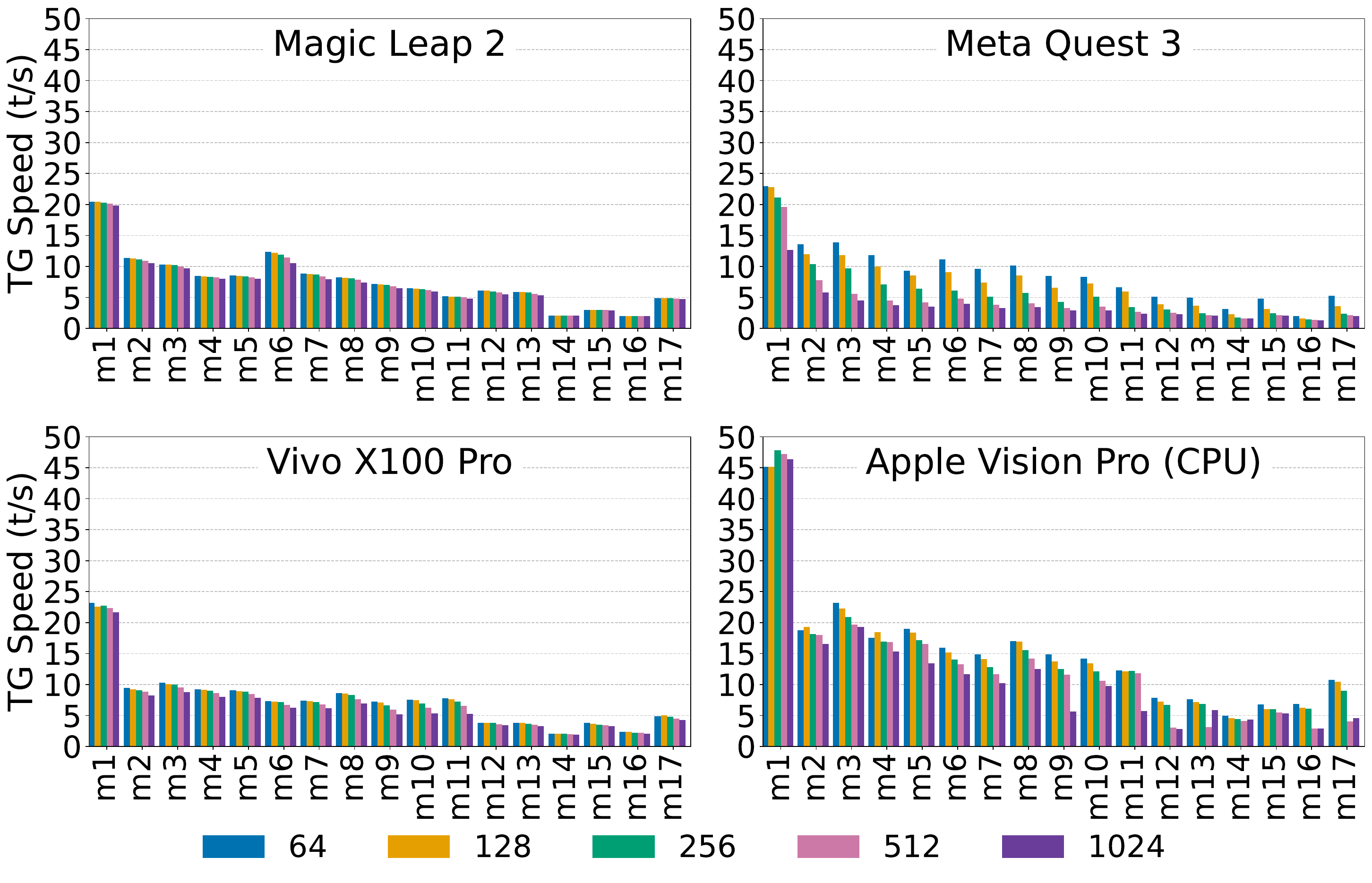}  
\vskip -0.45cm 
    \caption{Processing speed of the four devices in the TG test with varying string lengths: 64, 128, 256, 512, and 1024. The x-axis represents the model, while the y-axis represents the processing speed in t/s. Model IDs follow~\Cref{tab:models}:
Qwen (m$_1$),
Vikhr-Gemma (m$_2$–m$_5$),
Phi-3.1 (m$_6$–m$_{11}$),
LLaMA-2 (m$_{12}$–m$_{13}$),
Mistral-7B (m$_{14}$–m$_{17}$).}
    \label{fig:resTG:CPU}
\end{figure}

\begin{figure}
     \centering     
     \includegraphics[width=\linewidth]{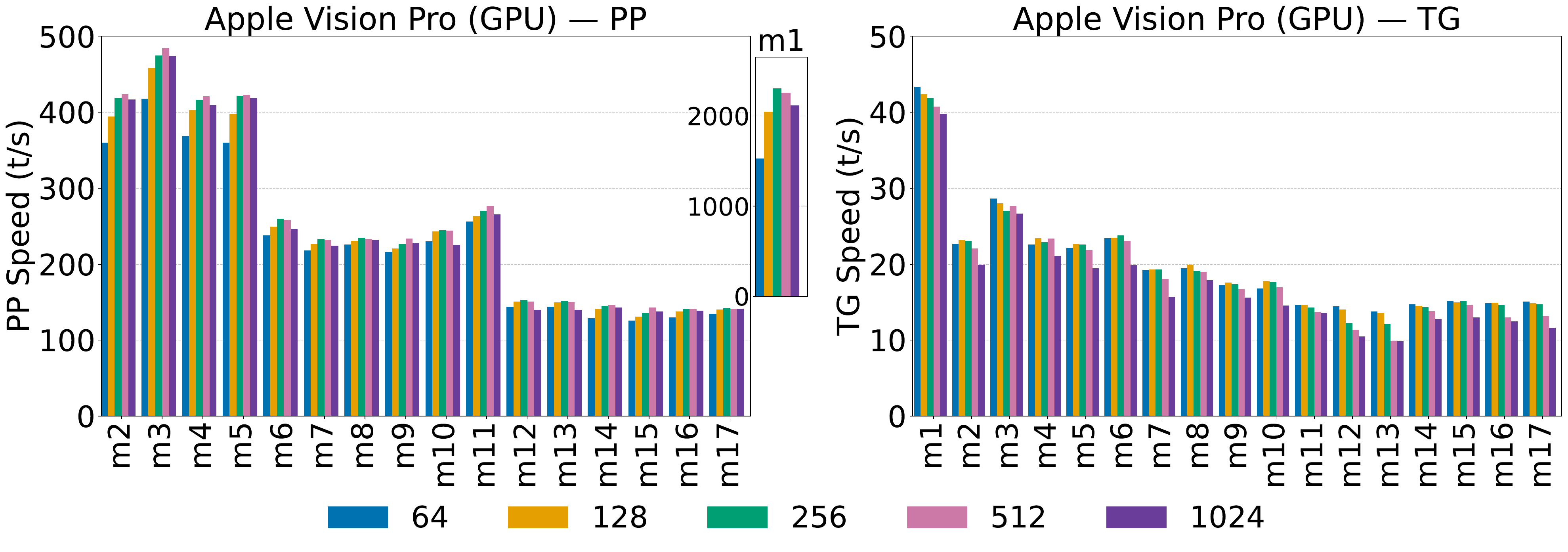}        
\vskip -0.45cm 
    \caption{GPU-based processing speed of the Apple Vision Pro for both PP (left) and TG (right) tests with varying string lengths: 64, 128, 256, 512, and 1024. Model IDs follow~\Cref{tab:models}: 
Vikhr-Gemma (m$_2$–m$_5$),
Phi-3.1 (m$_6$–m$_{11}$),
LLaMA-2 (m$_{12}$–m$_{13}$),
Mistral-7B (m$_{14}$–m$_{17}$).
}
    \label{fig:resPPTG:GPU}
\end{figure}
\subsection{Performance Consistency Results}
\label{sec:resultsStability}
\cref{fig:StabDD,fig:VPGPU} illustrate the consistency of performance across all model-device pairs. Across all 20 runs, the \gls{avp} demonstrates the most consistent results compared to other devices, with the corresponding speed also being the fastest. Notably, the GPU results for the \gls{avp} exhibit greater stability and smaller variance.
Among the remaining devices, the Magic Leap 2 also shows a reasonable degree of stability; however, for the first three models, its variance is surprisingly high, reaching 27\%, 32\%, and 26\% for \(m_2\), \(m_3\), and \(m_3\) under the PP setup, and 20\%, 23\%, and 27\% under the TG setup, respectively. In contrast, the Meta Quest 3 and Vivo X100 Pro exhibit relatively poor performance both in terms of speed and variance. Between these two devices, the Meta Quest 3 achieves higher speeds than the Vivo X100 Pro, but there is no clear difference in variance. For some models, one device shows lower variance, while for others, the variance is higher.
\subsection{Results of Processing Speed and String Length}
\label{sec:resultsPPTG} 
 \cref{fig:resPP:CPU,fig:resTG:CPU} present the processing speed results of PP and TG, respectively, across four CPU-based devices, with varying string lengths (prompt lengths and token sets). \cref{fig:resPPTG:GPU} presents the corresponding PP and TG results on GPU (AVP) , with varying string lengths (prompt lengths and token sets). Both PP and TG were evaluated with values of 64, 128, 256, 512, and 1024. The processing speed varies significantly across string lengths, devices, and models. \\
\noindent{}\textbf{String length and PP vs TG Speed}: PP speed is consistently higher than TG speed, with a speedup of approximately two to three times across the four (CPU-based) devices, while on \gls{avp} (GPU), the PP speed is significantly higher, ranging from 10 to 19 times the TG speed. Similarly, increasing the prompt size slightly reduces the prompt processing speed in PP, whereas in TG, the speed reduction is more significant with a growing token set. This suggests that TG has a higher computational cost due to its sequential execution of tokens, whereas PP is somewhat parallelized, requiring only the loading of prompts. Though there is minor variation in the 64, 128, and 256 string lengths, the last two, 512 and 1024, are consistently the slowest. When examining the variation across string lengths, \gls{mq3} exhibits the largest variance, whereas \gls{ml2} and \gls{avp} have the smallest variance.  \\
\begin{figure*}
    \centering 
    \includegraphics[width=0.998\linewidth]{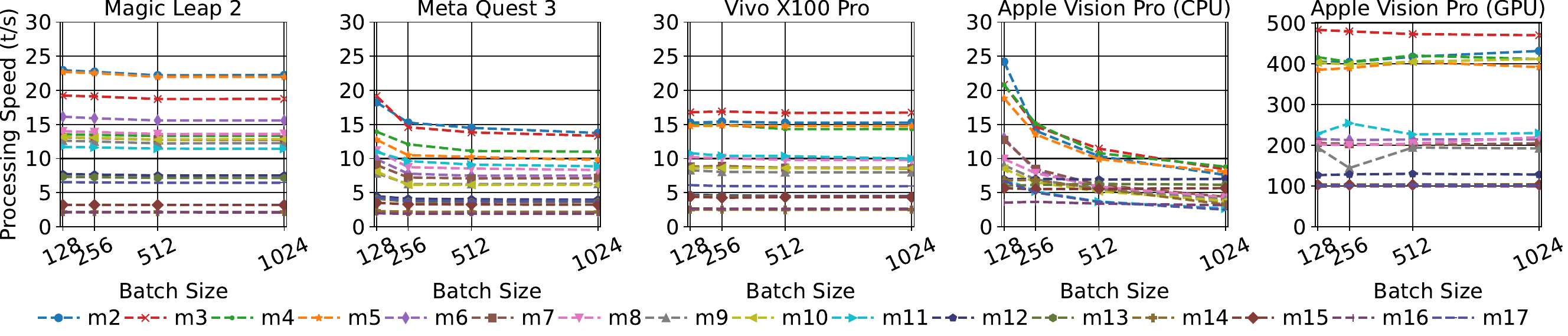}  
    \vskip -0.34780cm
  \caption{Batch Test results: Larger batch sizes significantly reduce throughput on CPU-based XR devices, while AVP–GPU remains stable. Model IDs follow~\Cref{tab:models}:
Vikhr-Gemma (m$_2$–m$_5$),
Phi-3.1 (m$_6$–m$_{11}$),
LLaMA-2 (m$_{12}$–m$_{13}$),
Mistral-7B (m$_{14}$–m$_{17}$).}
    \label{fig:resBT}
\end{figure*}
\begin{figure*}[!htbp]
    \centering  \includegraphics[width=0.998\linewidth]{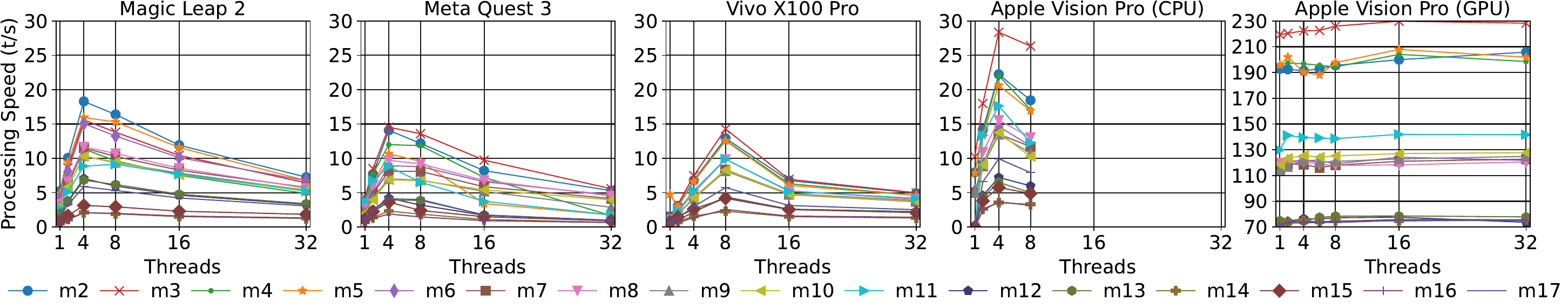}  
    \vskip -0.32478cm
\caption{Thread Test (TT) results: The X-axis represents the thread count, while the Y-axis shows the speed in tokens per second. Threads 4 and 8 deliver the fastest results. Note: \gls{avp} (CPU) fails for thread counts of 16 and 32. Model IDs follow~\Cref{tab:models}: 
Vikhr-Gemma (m$_2$–m$_5$),
Phi-3.1 (m$_6$–m$_{11}$),
LLaMA-2 (m$_{12}$–m$_{13}$),
Mistral-7B (m$_{14}$–m$_{17}$).}
    \label{fig:resThreadA}
\end{figure*}

\begin{figure*}[!htbp]
     \centering     
 \includegraphics[width=0.995\linewidth]{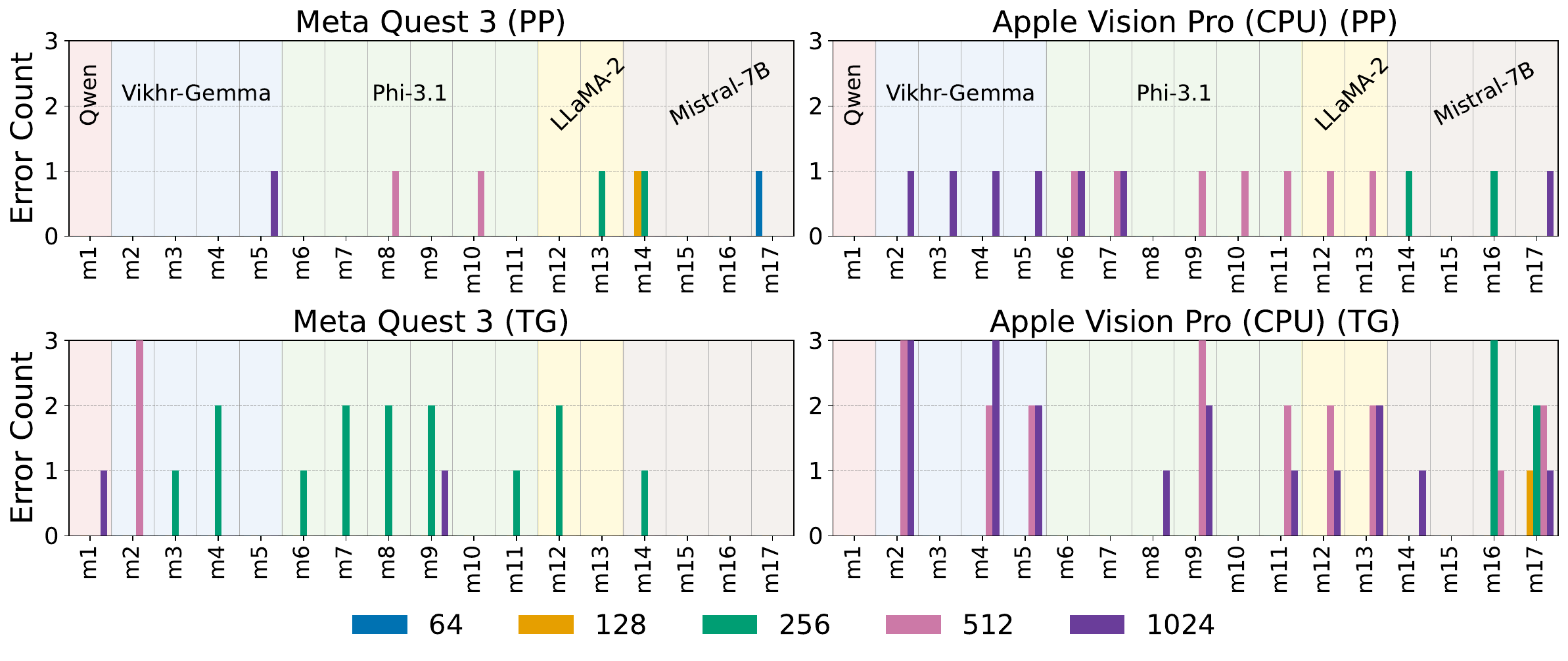}  
\vskip -0.355cm 
 \caption{Error counts for Meta Quest 3 (left) and \gls{avp} (right) in PP (top) and TG (bottom). Magic Leap 2, with zero errors, and Vivo X100 Pro, with a total of four errors, one in $m_1$ (PP-64), two in $m_{10}$ (PP-64, PP-128), and one in $m_{11}$ (TG-1024) are excluded. Model IDs follow~\Cref{tab:models}:
Qwen (m$_1$),
Vikhr-Gemma (m$_2$–m$_5$),
Phi-3.1 (m$_6$–m$_{11}$),
LLaMA-2 (m$_{12}$–m$_{13}$),
Mistral-7B (m$_{14}$–m$_{17}$).}
    \label{fig:resErrorsCount}
\end{figure*}

\noindent{}\textbf{Device-Based Analysis:} If we examine the devices, the overall performance trends reveal that \gls{avp} consistently shows the highest speed in both PP and TG tests. \gls{ml2} ranks second in both PP and TG performance, while Vivo X100s Pro outperforms \gls{mq3} in TG (particularly in longer strings 512, 1024) but ranks fourth in PP. This ranking highlights the varying computational capabilities of XR devices, with \gls{avp} (both CPU and GPU) standing out as the fastest one across all conditions. \gls{ml2} shows stable performance across all variations of string lengths, with lower $CV\%$ values, indicating that it does not vary significantly with changing string lengths. In contrast, \gls{mq3} shows a higher $CV\%$, indicating greater variance with increasing string length.  \\
\noindent{}\textbf{Error Counts:} \cref{fig:resErrorsCount} shows the errors counts during the PP and TG tests. Errors mostly occurred in \gls{avp} (CPU) and \gls{mq3}, particularly at longer prompt lengths of 512 and 1024 tokens. There are a few reasons for this. First, longer strings process more slowly, leading to abrupt changes and occasional failures. Second, memory constraints or inefficiencies in sustained generation contribute to these errors. With \gls{avp}, the general user experience is also suboptimal, as the device must be actively mounted on the head, which can cause errors if not handled carefully. Overall, \gls{mq3} records the highest count of errors, with frequent retries affecting its reliability, whereas \gls{ml2} remains highly stable, leading to minimal inconsistencies (zero errors). As expected, \gls{avp} (GPU) also reported zero errors. \\
\textbf{Model-Based Analysis:} In the model-wise analysis, smaller models such as Qwen2-0.5B and Vikhr-Gemma-2B achieve the highest speeds across devices, particularly in PP. In contrast, larger models like the LLaMA-2-7B series and Mistral-7B series exhibit significantly lower processing speeds, with higher variability and instability, especially in TG. Model size also impacts processing speed, as smaller models are faster. Here, $m_1$ is the fastest (though omitted from later analysis), while $m_2$, $m_3$, and $m_5$ are the fastest among the remaining models, whereas $m_{14}$ and $m_{16}$ are the slowest.

\subsection{Results of Parallelization with BT and TT}
\label{sec:resultsTTBT} 
This section presents the results of BT and TT, both of which are used to achieve concurrency and parallelization. Regardless of the parallelization method, the three models $m_2$, $m_3$, and $m_5$ consistently rank among the fastest across all devices, while $m_{14}$ and $m_{16}$ are the slowest.\\
\noindent\textbf{Results of BT:}  
\cref{fig:resBT} presents the results of the batch test with varying batch sizes of 128, 256, 512, and 1024. Generally, increasing the batch size leads to a decrease in performance across all four devices (except for \gls{avp} (GPU)) due to increased computational overhead. \gls{avp} (GPU), with its strong computational resources, does not show a significant performance drop with varying batch sizes, although there are some fluctuations at a batch size of 256. This indicates that GPU-based processing on \gls{avp} (GPU) is highly optimized for parallel execution and can effectively manage larger batches of input without affecting processing speed.

More importantly, \gls{avp} (CPU) experiences a noticeable performance drop as the batch size increases. However, the remaining devices do not follow this trend. For instance, on \gls{mq3}, batch size 128 yields the best performance, whereas for the other batch sizes, there is no significant change. Similarly, for \gls{ml2} and \gls{vivo}, variations in batch size do not appear to have a substantial impact on performance.\\
\noindent\textbf{Results of TT:}  
\cref{fig:resThreadA} presents the TT results for thread counts of 1, 2, 4, 8, 16, and 32. Across all four devices, processing speed shows the most significant increase when increasing the thread count from 1 to 4. The speed gain continues at 8 threads (or shows minor degradation), but beyond 8 threads, performance begins to decline slightly, with further degradation beyond 16 threads. \gls{avp} (CPU) fails at thread counts of 16 and 32, indicating its limitations in achieving this level of parallelism.

\gls{avp} (GPU) follows a different trend, maintaining consistently high processing speed across all thread counts. Unlike CPU-based devices, its performance remains stable even at 32 threads, demonstrating its superior ability to handle concurrent tasks. These results highlight that while the four CPU-based devices benefit from moderate threading, \gls{avp} (GPU) is significantly more efficient at scaling concurrency without experiencing notable performance degradation. Since our study primarily focuses on CPU-based implementation, we conclude that using a moderate thread count of 4, 6, or 8 yields optimal results.
\begin{figure}
    \centering
    \includegraphics[width=1\linewidth]{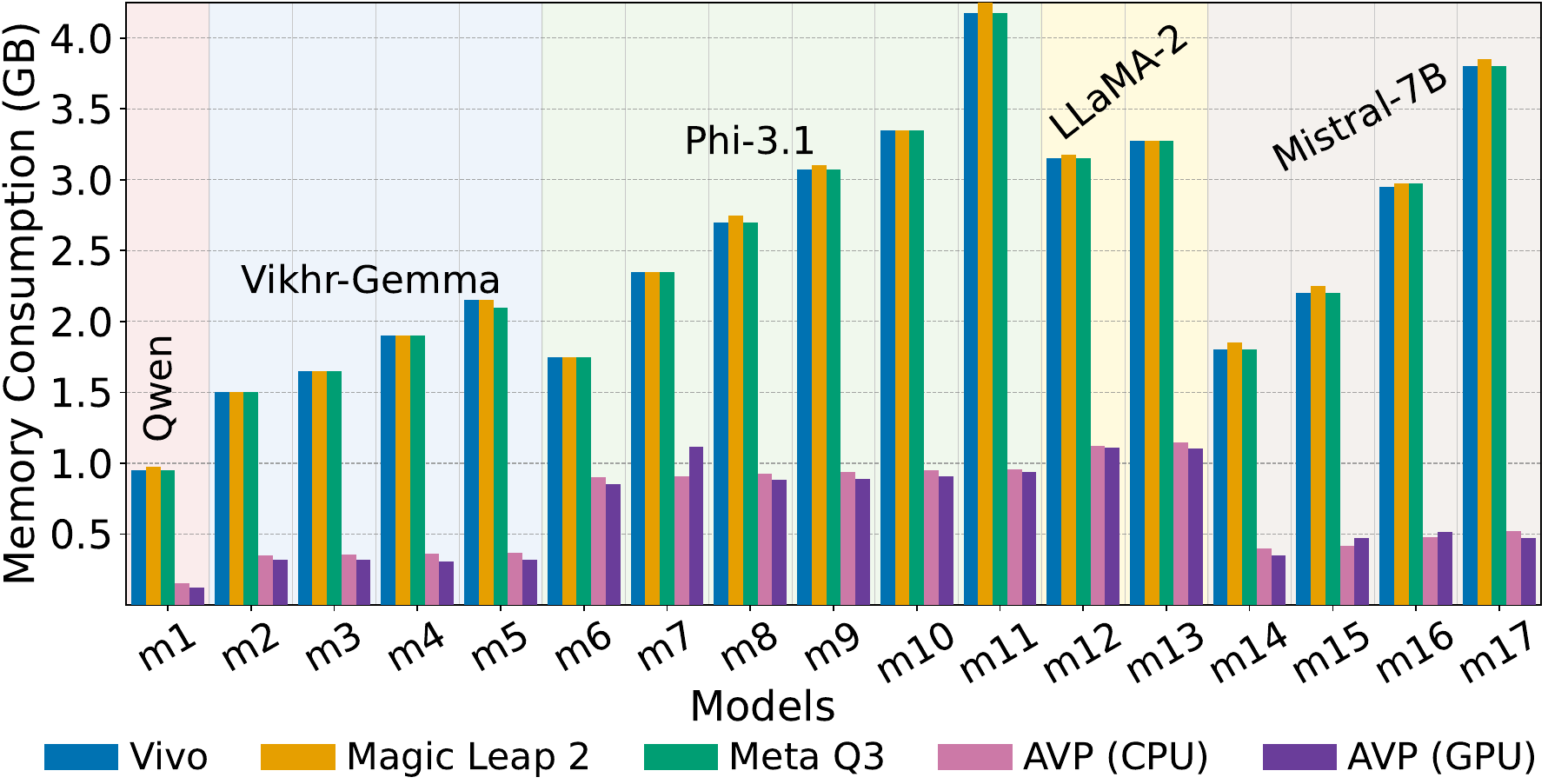}
    \vskip -0.196478cm 
    \caption{Memory consumption for each model-device pair. The results represent the mean values across batch sizes 128, 256, 512, and 1024.}
    \label{fig:Memory_plot}
    \vskip -0.396478cm 
\end{figure}
\subsection{Memory Consumption Results }
\label{sec:resultsMemory}
\cref{fig:Memory_plot} presents the memory consumption results for each model-device pair. The values represent the mean memory consumption from five experiments with varying batch sizes of 128, 256, 512, and 1024. As expected, memory consumption varies across models, with most showing consistent usage across different devices.    Generally, within each model series, memory consumption increases with model size. However, minor exceptions exist, such as $m_7$ on \gls{avp} (GPU), which consumes more memory than other models in the Phi-3.1 series $[m_6-m_{11}]$. This size-based trend does not necessarily hold across series, even for models of similar sizes. For example, \textit{m2} (1.36 GB) consumes less memory than  $m_6$ (1.32 GB) despite having a larger size, as they belong to different series. Similarly,  $m_7$ (1.94 GB) requires more memory than \textit{$m_5$} (2.0 GB) and \textit{$m_{15}$} (2.05 GB), as all three belong to different series. Another example is  $m_{12}$ (2.63 GB) consuming more memory than $m_{16}$ (2.81 GB), despite being smaller in size.

From a device-specific perspective, \gls{avp} demonstrates significantly lower memory consumption than the other devices, averaging $0.5782$ GB on the CPU and $0.5488$ GB on the GPU across all $17$ models. This efficiency is attributed to Apple Silicon’s unified memory architecture and the optimized memory management of the \gls{avp}, which can load only the active portions of a model during execution; for further details, see Feng et al.~\cite{feng2025profilingapplesiliconperformance}.

Among the remaining devices, Meta Quest 3 consumes the least memory at $2.3520$ GB, followed by Vivo X100 Pro at $2.3540$ GB, while Magic Leap 2 exhibits the highest at $2.3748$ GB. These findings suggest that \gls{avp} (both CPU and GPU) is approximately four times more memory-efficient than Magic Leap 2, Meta Quest 3, and Vivo X100 Pro. Its superior memory management and hardware optimizations make it the most efficient device in terms of memory consumption.

\subsection{Battery Consumption Results}
\label{sec:resultsBattery}
\begin{figure*}[!htbp]
    \centering   
    \includegraphics[width=0.49675\linewidth]{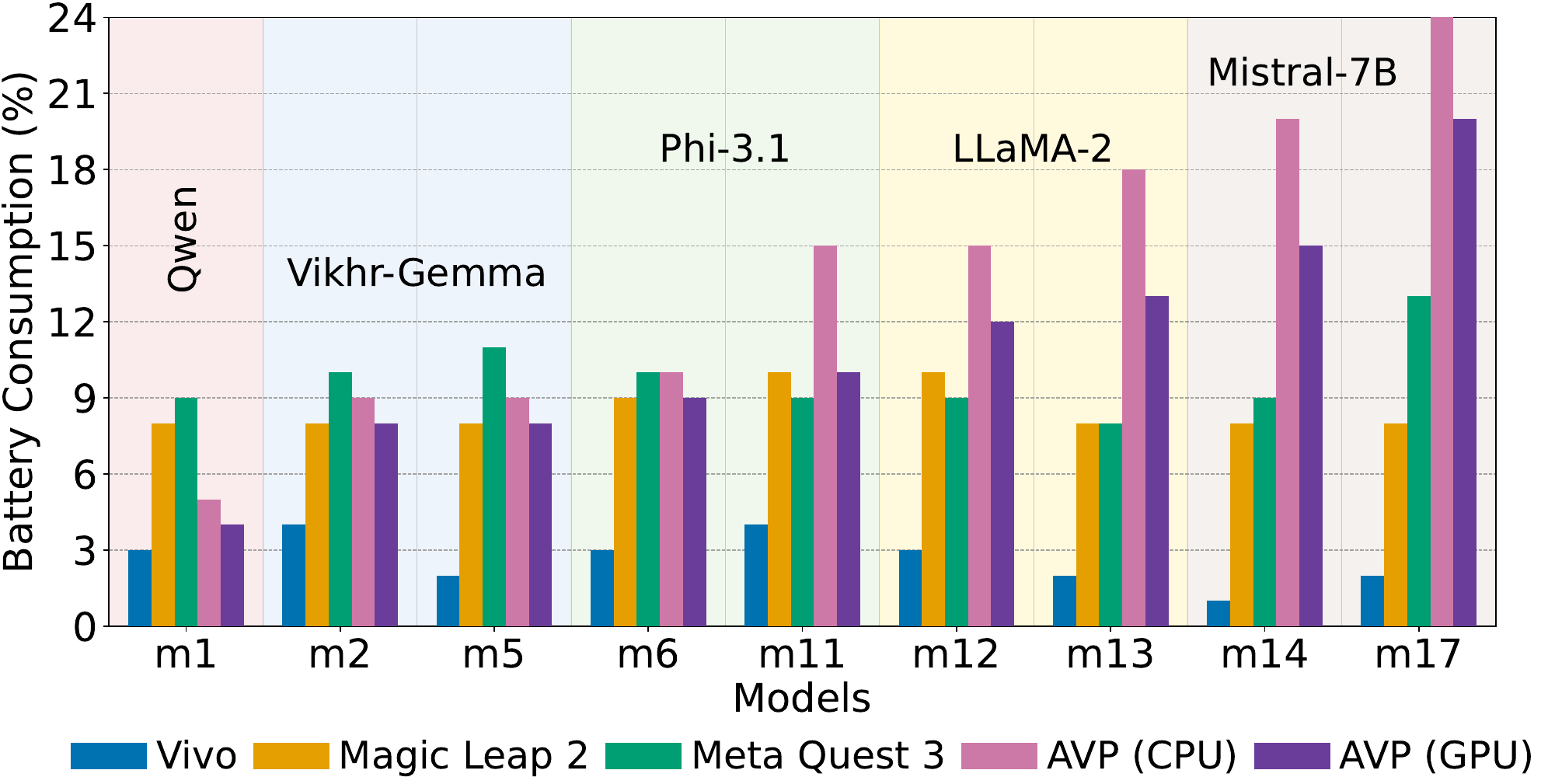}  \hfill 
    \includegraphics[width=0.49675\linewidth]{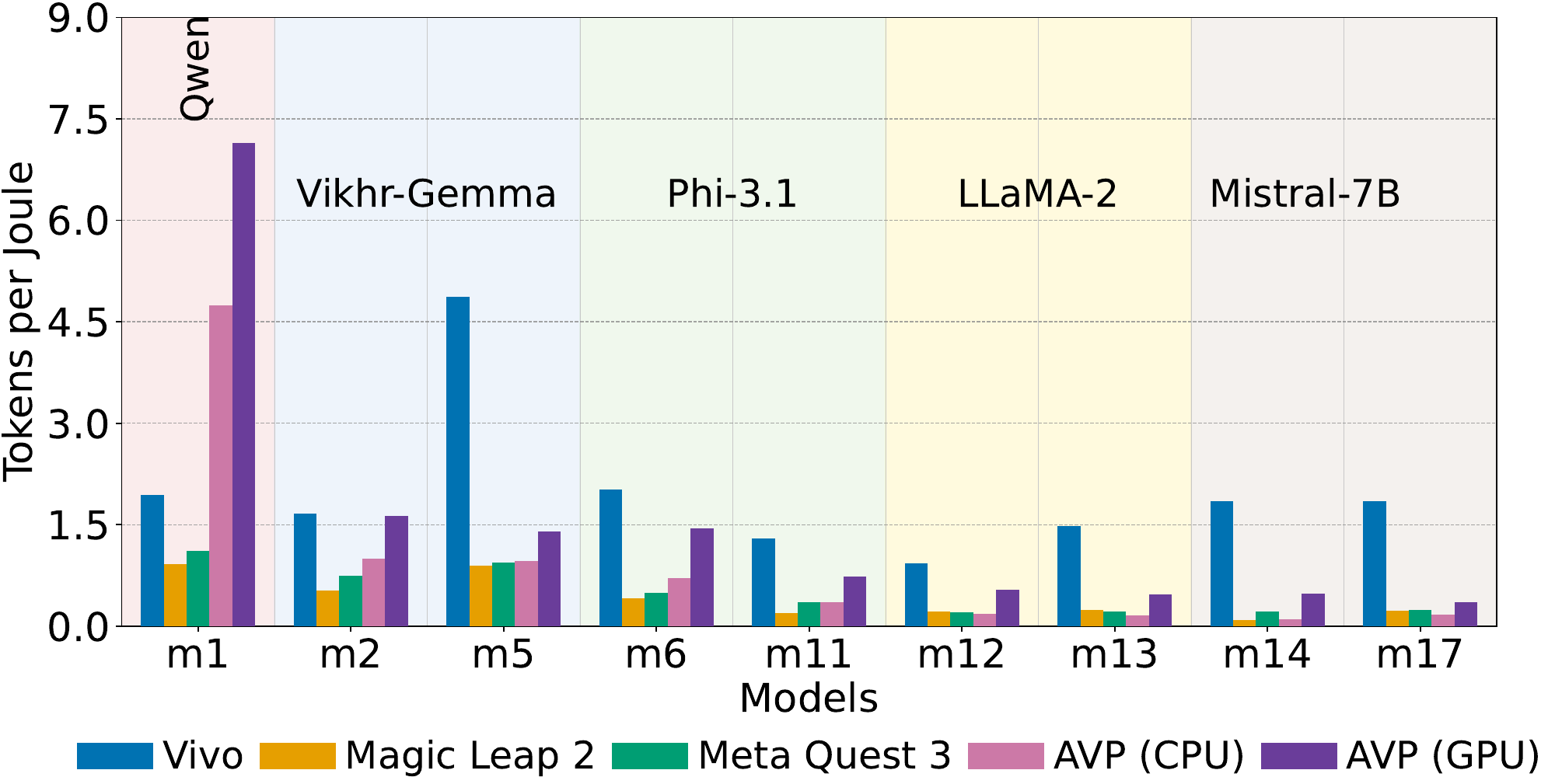}  
    \vskip -0.396478cm 
    \caption{Battery consumption and energy efficiency across models and devices. 
\textbf{Left:} Battery consumption (\%, \textbf{$\downarrow$}) over a 10-minute continuous inference run. 
\textbf{Right:} Tokens-per-joule (\textbf{$\uparrow$}) computed by dividing total generated tokens by the estimated energy usage during the same interval. 
Results are averaged over three runs.}

    \label{fig:BatteryTEST}  
\end{figure*}

\cref{fig:BatteryTEST} (left) presents the average battery consumption results over a 10-minute experiment, with each test repeated three times. We observe that for the first two series, the Qwen Series and Vikhr-Gemma Series, battery consumption remains relatively low and with lower variation across all four devices. However, for larger models in the LLaMA-2 Series and Mistral-7B Series, battery consumption increases significantly. 
\gls{avp} shows a strong correlation between model size and battery consumption, both for GPU and CPU usage. In contrast, the remaining three devices do not exhibit a significant variation with respect to model size. \gls{vivo} demonstrates the best battery life, with an average loss of only $2.5\%$ over 10 minutes. \gls{ml2} follows with a $8.5\%$ average battery loss, while \gls{mq3} records a slightly higher battery loss at $9.7\%$. \gls{avp} shows the worst in this regard, with GPU usage resulting in a $10.1\%$ battery loss and with CPU usage leading to a $12.6\%$ battery loss in 10 minuets. 
\cref{fig:BatteryTEST}~(right) reports the tokens-per-joule metric, computed by dividing the total number of generated tokens by the estimated energy usage during the same 10-minute interval. Higher values indicate better energy efficiency.  
Overall, smaller models exhibit noticeably higher energy efficiency, while larger models consume more energy per token.  
Across devices, the Vivo X100 Pro achieves the best energy efficiency, followed by the AVP (GPU and CPU), with the MQ3 and ML2 showing comparatively lower efficiency.

\begin{figure}[!htbp]
    \centering  
    \includegraphics[width=1.0\linewidth]{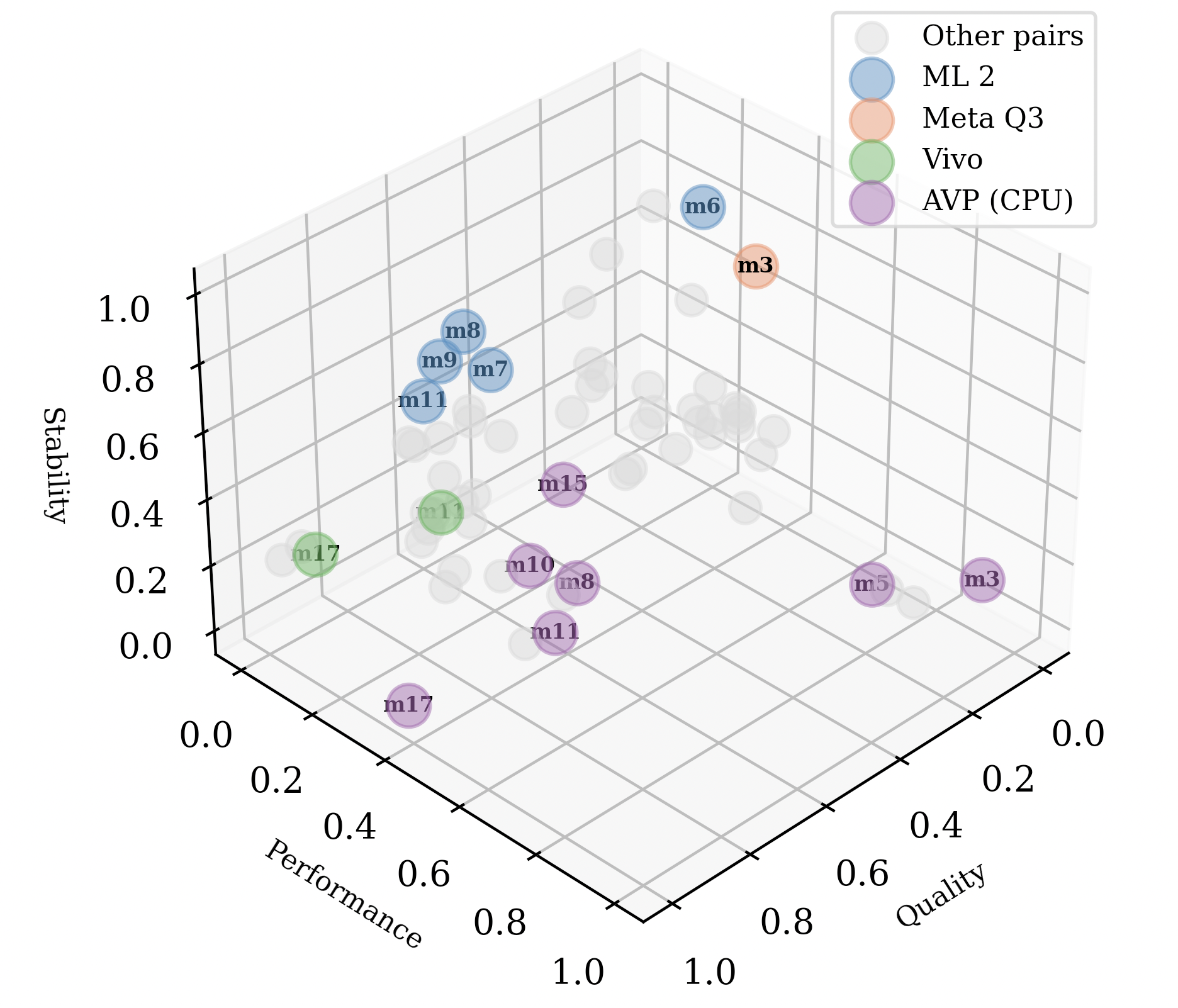}
    \vskip -0.478cm 
    \caption{Pareto fronts for various device-model pairs, with different colors representing distinct device types. Performance is evaluated based on processing speed, memory consumption, and battery consumption. Stability is assessed using the \textit{CV} and \textit{error count} values of \textit{PP} and \textit{TG}. Quality is measured through evaluation results on six benchmark datasets.
    } 
    \label{fig:pareto}
\end{figure}

\subsection{Pareto Front Results}
Since we use both CPU and GPU for inference in the Vision Pro experiments, while the other three devices use CPU only for inference, we exclude \gls{avp} GPU from the Pareto calculation to ensure a fair comparison. With 4 devices and 16 models (m2-m17), we have a total of 64 device-model pairs. Using \cref{equ:pareto} and \cref{equ:score}, we calculated the \textit{quality}, \textit{stability} and \textit{performance}  scores for each pair, then identified the Pareto fronts and visualized them in \cref{fig:pareto}. The quality of a model can be evaluated using various benchmarks, such as accuracy and perplexity, as detailed in \cref{sec:modelsQA}.  For the performance objective, we evaluate performance using \textit{PP}, \textit{TG}, \textit{memory consumption}, and \textit{battery consumption}, with respective weights of 0.35, 0.35, 0.2, and 0.1. For the stability objective, we assess it based on the coefficient of variation (\textit{CV}) and error count of \textit{PP} and \textit{TG}, assigning 0.7 to CV and 0.3 to error count.

\cref{fig:pareto} illustrates the Pareto front points, highlighting the optimal choices across different model sizes and devices. From a device-level perspective, \gls{avp} (CPU) achieves the highest speed among the evaluated platforms, making it suitable for latency-sensitive deployments. However, this speed advantage comes at the cost of reduced stability, indicating potential variability in output quality under prolonged or complex workloads. In contrast, ML 2 consistently provides the greatest stability, suggesting its suitability for applications where robustness and predictable behavior are critical. Meanwhile, both Vivo and Meta Q3 exhibit a more balanced trade-off between speed and stability, positioning them as general-purpose deployment options when neither metric can be sacrificed significantly.

At the model level, several configurations emerge as context-dependent optimal choices. Specifically, $m_{11}$ and $m_{17}$ provide high quality. Meanwhile, $m_3$ is selected as a Pareto-optimal choice due to its high performance on \gls{avp} (CPU) and robust stability on Meta Q3, making it a compelling option for cross-device consistency when applications require portability across heterogeneous environments. A similar trend is observed with $m_8$, which attains strong performance on \gls{avp} (CPU), while maintaining commendable stability on ML 2. This suggests that certain mid-sized models can achieve favorable multi-objective trade-offs across both compute-constrained and stability-oriented devices.

Overall, the Pareto analysis reveals that optimal configurations are inherently deployment-dependent. Rather than a single universally superior model–device pair, our results indicate distinct operating regimes across different CPU-based platforms: speed-dominant settings optimized for low-latency execution, stability-oriented configurations emphasizing consistent behavior, and balanced trade-off platforms that maintain reasonable performance across both objectives. 

Although all evaluations are conducted under CPU inference, variations across devices still lead to meaningful differences in efficiency and robustness. These findings provide actionable guidance for practitioners selecting model-device combinations under concrete resource and application constraints.

%% file: content/05-Eval-Analysis.tex
\section{Analysis of On-Device LLMs}
\label{sec:analysis}
In this section, we evaluate the applicability of on-device LLMs in two interactive applications and report performance in terms of accuracy. Additionally, to assess efficiency, we compare on-device LLMs with alternative deployment approaches, namely client-server and cloud-based setups.
\subsection{Evaluation on Interactive Applications}
We conducted an analysis of our on-device LLM setup (\aivaluatexr) using datasets from two existing LLM-based interactive systems: VOICE~\cite{Jia2024_1} and GeoVis~\cite{mena2025augmentingLLMs}. VOICE employs LLMs for interactive exploratory visualization and provides natural language interaction capabilities that allow users to navigate and manipulate 3D biological models in real time. GeoVis is another LLM-based method applied to geospatial data, enabling tasks such as selecting the viewing direction or retrieving location information through conversational queries. 
The details of the metrics, prompts and queries used are provided in the supplementary material. For this evaluation, we used the following metrics:
\begin{itemize}
    \item \textit{Formatting (GeoVis):} Measures how well the model's output conforms to the expected \texttt{[latitude, longitude]} format. A score of 100\% is assigned if the formatting is perfect, 75\% if it is off by one character, 50\% if it is otherwise incorrect but still easy to parse, and 0\% if it cannot be parsed.
    \item \textit{Accuracy (GeoVis):} Evaluates the spatial accuracy of the predicted coordinates relative to the target location. The score is 100\% if the coordinates are near the center of the target area, 20\% if they are off target but the area of interest remains partially visible at a reasonable zoom level, and 0\% if it is not visible at all.
    \item \textit{Score:} For the VOICE dataset, each query is graded 100\% if the output is correct and 0\% otherwise. For the GeoVis dataset, if the Formatting score is 0, the overall Score is also 0. Otherwise, it is computed as a weighted combination of Formatting (20\%) and Accuracy (80\%) to reflect both structural and spatial correctness.
\end{itemize}

Since our current study focuses specifically on LLMs, we did not take evaluate the LLMs on prompts requiring a vision model or relying on fine-tuned models for modeling tasks. Instead, we used the VOICE and GeoVis datasets and evaluated only the queries that LLMs could answer on their own, which is sufficient for evaluating their accuracy.
   \Cref{tab:geovis_voice_grouped} presents the accuracy results. Since these tasks involve complex interactions, we observe that some models completely fail in terms of accuracy.

\begin{table}[!tb]
 \caption{
Performance summary of models \(m_8\) to \(m_{17}\) on the 
\textit{GeoVis}~\cite{mena2025augmentingLLMs} and 
\textit{VOICE}~\cite{Jia2024_1} datasets.
Values are reported as $\mu \pm \sigma$, where the notation is descriptive and does not imply a symmetric interval around the mean.
The Score metric is computed as a weighted sum of Formatting (20\%) and Accuracy (80\%).
Model IDs follow~\Cref{tab:models}: 
Phi-3.1 (m$_8$–m$_{11}$),
LLaMA-2 (m$_{12}$–m$_{13}$),
Mistral-7B (m$_{14}$–m$_{17}$).
}
\vskip -0.126478cm 
\centering
\renewcommand{\arraystretch}{1.1}
\resizebox{\linewidth}{!}{
\begin{tabular}{c|ccc|c}
\hline 
Model ID& \multicolumn{3}{c|}{GeoVis data} & VOICE data \\
\cline{2-4}
 & Formatting ($\mu \pm \sigma$) & Accuracy ($\mu \pm \sigma$) & Score ($\mu \pm \sigma$) &  Accuracy ($\mu \pm \sigma$) \\
\hline 
$m_8$  & 40.00 $\pm$ 20.34 & 55.00 $\pm$ 44.70 & 52.00 $\pm$ 38.44 & 0.00 $\pm$ 0.00 \\
$m_9$  & 40.00 $\pm$ 20.34 & 57.33 $\pm$ 46.23 & 51.20 $\pm$ 39.85 & 0.00 $\pm$ 0.00 \\
$m_{10}$ & 36.67 $\pm$ 22.49 & 71.00 $\pm$ 41.63 & 56.13 $\pm$ 40.13 & 3.70 $\pm$ 19.25 \\
$m_{11}$ & 45.00 $\pm$ 15.26 & 64.00 $\pm$ 45.00 & 60.20 $\pm$ 37.57 & 0.00 $\pm$ 0.00 \\
$m_{12}$ & 0.00 $\pm$ 0.00  & 0.00 $\pm$ 0.00  & 0.00 $\pm$ 0.00  & 3.70 $\pm$ 19.25 \\
$m_{13}$ & 1.67 $\pm$ 9.13  & 0.00 $\pm$ 0.00  & 0.33 $\pm$ 1.83  & 0.00 $\pm$ 0.00 \\
$m_{14}$ & 38.33 $\pm$ 21.51 & 6.00 $\pm$ 15.89 & 11.93 $\pm$ 14.03 & 7.41 $\pm$ 26.69 \\
$m_{15}$ & 52.50 $\pm$ 7.63  & 39.33 $\pm$ 42.74 & 41.97 $\pm$ 34.92 & 48.15 $\pm$ 50.92 \\
$m_{16}$ & 13.33 $\pm$ 26.04 & 17.33 $\pm$ 36.29 & 16.53 $\pm$ 33.70 & 40.74 $\pm$ 50.07 \\
$m_{17}$ & 54.17 $\pm$ 13.27 & 87.33 $\pm$ 29.47 & 80.70 $\pm$ 23.63 & 51.85 $\pm$ 50.92 \\
\hline
\textit{GPT-4o} & 100 $\pm$ 0 & 99.66 $\pm$ 1.82 & 99.73 $\pm$ 1.43& 96.30 $\pm$ 19.25 \\
\textit{GPT-4o-Mini} & 98.33 $\pm$ 9.13& 93.33 $\pm$ 22.94 & 94.33 $\pm$ 19.79& 85.19  $\pm$ 36.20 \\
\hline
\end{tabular}}
\label{tab:geovis_voice_grouped}
\end{table}
\begin{table}[!tb]
\centering
\caption{Time comparison (in seconds), including both model loading (warm-up) time and first response latency, for model \(M_{15}\) running locally on XR devices. Each cell reports the mean and standard deviation \((\mu \pm \sigma)\).}
\vskip -0.2478cm 
\resizebox{\linewidth}{!}{
\begin{tabular}{c|c|c|c|c|c}
\hline
Model ID & Series & AVP & ML2 & Vivo& MQ3 \\
\hline
  $m_{15}$   &  Mistral-7B & 2.22 $\pm$ 1.05 & 61.21 $\pm$ 38.04 & 86.69 $\pm$ 53.84 & 82.96$\pm$53.36\\
\hline
\end{tabular}
}
\label{tab:m15_time_comparison}

\end{table}
\begin{table*}[!tb]
\centering
\caption{Comparison of cold and warm first response time (F.R.T) in seconds for seven LLMs accessed via AVP across three deployment modes.  Each cell reports the mean and standard deviation \((\mu \pm \sigma)\). Here, $\times$ indicates that the model is not supported in that mode.}
\vskip -0.2cm
\resizebox{\linewidth}{!}{
\begin{tabular}{|c|c|c|c|c|c|c|c|c|}
\hline 
Type & Approach & \makecell{GPT-4o} & \makecell{GPT-4o\\Mini} & \makecell{Qwen2\\($m_1$)} & \makecell{Vikhr-Gemma\\($m_5$)} & \makecell{Phi-3.1\\($m_{11}$)} & \makecell{LLaMA-2\\($m_{13}$)} & \makecell{Mistral\\($m_{17}$)} \\
\hline
\multirow{4}{*}{Cold FRT}
  & Client-Server (CPU)   & $\times$ & $\times$ & 0.67 $\pm$ 0.66 & 1.94 $\pm$ 1.36 & 2.22 $\pm$ 1.05 & 2.65 $\pm$ 1.54 & 3.72 $\pm$ 0.71 \\
  
  &Client-Server (GPU)   & $\times$ & $\times$ & 1.16 $\pm$ 0.25 & 1.73 $\pm$ 0.33 & 1.43 $\pm$ 0.32 & 1.84 $\pm$ 0.61 & 2.72 $\pm$ 0.64 \\
  
  & On-Device  (AVP)              & $\times$ & $\times$ & 1.07 $\pm$ 0.06 & 2.56 $\pm$ 0.37 & 3.06 $\pm$ 1.17 & 3.30 $\pm$ 0.93 & 3.86 $\pm$ 0.68 \\
  
  &MacBook (GPU)    & $\times$ & $\times$ & 1.12 $\pm$ 0.62 & 1.83 $\pm$ 0.17 & 1.29 $\pm$ 0.11 & 1.54 $\pm$ 0.19 & 2.21 $\pm$ 1.10\\
\hline
\multirow{5}{*}{Warm F.R.T}
  & Cloud-based           & 1.13 $\pm$ 0.29 & 1.03 $\pm$ 0.31 & $\times$ & $\times$ & $\times$ & $\times$ & $\times$ \\
  & Client-Server (CPU)   & $\times$ & $\times$ & 0.25 $\pm$ 0.15 & 1.36 $\pm$ 0.69 & 1.37 $\pm$ 0.07 & 1.40 $\pm$ 0.23 & 2.91 $\pm$ 0.68 \\
  
  & Client-Server (GPU)    & $\times$ & $\times$ & 0.32 $\pm$ 0.11 & 0.47 $\pm$ 0.04 & 0.37 $\pm$ 0.04 & 0.50 $\pm$ 0.06 & 0.65 $\pm$ 0.03 \\
  
  & On-Device (AVP)           & $\times$ & $\times$ & 0.03 $\pm$ 0.00 & 0.26 $\pm$ 0.28 & 0.12 $\pm$ 0.04 & 0.27 $\pm$ 0.28 & 0.07 $\pm$ 0.00 \\
  
  & MacBook (GPU)    & $\times$ & $\times$ & 0.16 $\pm$ 0.12 & 0.20 $\pm$ 0.31 & 0.18 $\pm$ 0.29 & 0.22 $\pm$ 0.07 &0.23 $\pm$ 0.02\\
\hline

\hline
\end{tabular}
}
\label{tab:latency7LLMs}
\end{table*}
\subsection{Latency Comparison Across Deployment Modes}
Latency, or first response time (F.R.T), refers to the duration between issuing a prompt and receiving the first token of the response, whereas warm-up time denotes the initial time required to load the model into memory. \textit{Cold F.R.T} refers to the latency observed on the first prompt after model initialization (i.e., warm-up $+$ response), while \textit{Warm F.R.T} reflects the prompt-to-first-token latency for subsequent queries on an already loaded model.
We compared latency across three deployment modes: cloud-based, client-server, and on-device. In the client-server setup, the LLMs were hosted on a \textit{MacBook Pro} with an \textit{M4 Pro chip} and 48\,GB of \textit{RAM}, and accessed via AVP. To begin, we evaluated model $m_{15}$ across all four XR devices (see~\cref{tab:m15_time_comparison}) and selected the AVP as the fastest device for subsequent comparisons.  
We include~\cref{tab:m15_time_comparison} as an additional device-level check using a representative large model ($m_{15}$), confirming that the AVP provides the lowest latency among the XR devices and is therefore the appropriate reference platform for the detailed F.R.T. analysis reported in~\cref{tab:latency7LLMs}.
For a broader evaluation, we selected five representative models (one per category) and benchmarked them across all three deployment modes. Each experiment was repeated three times, and we report the mean latency along with standard deviations. 

As shown in~\cref{tab:latency7LLMs}, these five models were evaluated alongside GPT-4o and GPT-4o-mini, all accessed via the AVP. While on-device models exhibit longer warm-up times (\ie higher \textit{Cold F.R.T}), they consistently achieve the lowest latency after loading (\ie lower \textit{Warm F.R.T}), making them well-suited for real-time XR interactions, since warm-up only happens once during a session. Importantly, the latency values in~\cref{tab:latency7LLMs} correspond to a simple prompt--\textit{``Give me 20 continuous words as text only''}. 

For longer or more complex prompts, latency will naturally increase. Longer prompts require processing a larger number of input tokens, which raises inference time. For cloud-based models, this effect is further amplified because the entire prompt must be transmitted over the network, adding additional round-trip delay. Thus, the latency reported in~\cref{tab:latency7LLMs} should be interpreted as a lower bound for cloud-based deployments.
On-device models also experience increased latency with longer prompts, but the growth is more predictable because it depends solely on local computation. This explains why delays can increase in real scenarios, particularly for cloud-based models.\\
\cref{tab:latency7LLMs} also includes GPU-based latency measurements from both a client–server setup and a MacBook Pro. As expected, GPU acceleration reduces warm latency for the larger models (e.g., Phi-3.1, Mistral), which can exploit the parallelism. In contrast, small models such as Qwen2 (m$_1$) show little or no benefit, since the GPU transfer overhead outweighs the compute savings. These observations match typical desktop/server inference trends and clarify that several latency behaviors observed on XR devices, particularly those shaped by CPU-only execution and thermal constraints are specific to XR platforms.\\
Our results show that several models achieve relatively low accuracy on the XR datasets. This is expected, as none of the evaluated LLMs were fine-tuned on XR-specific data. Fine-tuning would likely give substantial accuracy improvements. Likewise, some models exhibit lower computational efficiency, but this can be addressed through standard techniques such as model compression and quantization, which reduce latency and resource usage. Taken together, these considerations reinforce that \aivaluatexr, \ie our attempt at deploying LLMs fully on-device will serve as a foundational baseline for future research in this emerging area.
We also clarify that for GPT-4o and GPT-4o-mini, all prompts were sent from a local computer to OpenAI’s servers, and responses were received over the Internet. This inherently introduces network latency, consistent with observations reported in XaiR~\cite{Srinidhi2024ISMAR:XaiR}. If an XR device communicates directly with the cloud (rather than a PC acting as a relay), an additional client-server delay may occur, which can further increase end-to-end latency.

To contextualize our findings, we note that the efficiency of on-device LLMs is expected to improve as XR hardware evolves. Future devices will likely incorporate more powerful processors and dedicated AI accelerators, which will reduce on-device inference time. Although cloud-based LLMs will also become faster, some degree of network latency will always persist and is unlikely to shrink substantially. In our experiments, the relatively small difference in latency between GPT-4o and GPT-4o-mini suggests that network delay already dominates their overall response time~\cite{gustafson2011amdahl}. Therefore, we conclude that the first-response advantage of on-device LLMs is likely to increase over time, while their warm-up cost will diminish as XR hardware improves and as LLMs undergo targeted fine-tuning and compression.

%% file: content/06-Discussion.tex
\section{Discussion}
\label{sec:discussion}
\aivaluatexr provides a systematic framework for evaluating LLMs running locally on XR devices, taking into account both model- and device-specific factors.
Using \aivaluatexr, we have performed extensive testing of multiple open-source language models on four distinct XR devices. Our study identifies critical performance evaluation factors and introduces a suite of testing scripts to analyze these aspects systematically. These scripts enhance the reproducibility of our findings, making them relevant for future analyses in similar contexts.  The processing speed, measured in tokens per second, was evaluated across multiple dimensions. This includes a consistency analysis to determine whether each model-device pair maintains stable performance over time or exhibits fluctuations. The impact of string length on both \gls{pp} and \gls{tg} was analyzed for each model-device combination. Additionally, the study examined the support for parallelization and concurrency control through BT and TT tests. Furthermore, memory usage and battery consumption were also assessed to provide a holistic evaluation of system performance.
Our goal in this section is not only to summarize empirical trends but also to distill practical insights for XR developers, system designers, and researchers deciding which model-device configurations are suitable for real-time deployment.\\ 
Beyond general LLM trends, several findings from our experiments are specific to XR deployment constraints.
First, XR devices exhibit strong \textit{thermal sensitivity}, which significantly impacts performance consistency over repeated runs, \ie an effect far less pronounced on desktops or smartphones.
Second, because most XR devices rely primarily on \textit{CPU-only} inference, parallelization behaves differently than on GPU-based systems: increasing thread count beyond a moderate range yields diminishing returns and may even reduce throughput due to thermal throttling.
Third, our \textit{battery consumption and energy-efficiency} analysis highlights practical limits for prolonged XR usage, where even small differences in model size translate into meaningful differences in sustained runtimes.
Finally, our evaluation exposes \textit{deployability constraints}, such as memory ceilings that restrict which model sizes can be executed on specific XR devices.
These XR-specific effects illustrate that behavior commonly reported in standard LLM benchmarks does not directly generalize to XR platforms, underscoring the need for a dedicated XR-focused evaluation framework such as \aivaluatexr.

\subsection{Summary of Results}
\cref{fig:VPGPU,fig:StabDD} present the consistency results in ascending order, where we see that the \gls{avp} GPU exhibits the highest stability.  
While \gls{ml2} shows a similar level of stability to the \gls{avp} CPU, its first three models ($m_2$, $m_3$, and $m_4$) exhibit a CV above $20\%$ for TG and above $26\%$ for PP.
Considering the error count in~\cref{fig:resErrorsCount}, \gls{ml2} emerges as the most stable device, having recorded zero errors. In contrast, the \gls{avp} CPU encountered multiple errors. The remaining two devices, \gls{vivo} and \gls{mq3}, exhibit higher instability, with \gls{vivo} performing relatively better than \gls{mq3}. This also indicates that consistency is model-dependent, as seen with models $m_2$, $m_3$, and $m_4$, which exhibited reduced stability.

For $m_1$, being the smallest model, all devices performed efficiently. Among the remaining models ($m_2$ to $m_{16}$), the average inference speeds  across various devices (according to PP and TG tests) are as follows. \gls{avp} (CPU) achieved the highest inference speed  (19.91 t/s PP, 11.54 t/s TG), followed by \gls{ml2} (12.18 t/s PP, 6.73 t/s TG), \gls{mq3} (10.89 t/s PP, 5.17 t/s TG), and \gls{vivo} (9.35 t/s PP, 6.12 t/s TG), reflecting performance variations across XR devices. These speeds meet basic conversational requirements and enable LLM inference for broader applications beyond dialogue-based tasks. The PP speed across all devices is  two to three times faster than TG speed, while for \gls{avp} GPU PP speed is 10 to 19 time faster than TG speed. This is because PP only encodes the input, whereas TG must process sequential dependencies during decoding. 
In model-based analysis, smaller models such as $m_1$, $m_2$, $m_3$, and $m_5$ are the fastest, while $m_{14}$ and $m_{16}$ were the slowest. Regarding string length, larger PP and TG inputs (512, 1024) are slower than shorter ones (64, 128, 256). Moreover, longer strings (512, 1024) frequently result in errors. Among devices, \gls{mq3} exhibited the highest number of errors, followed by the \gls{avp} CPU and then \gls{vivo}. In contrast, \gls{ml2} and the \gls{avp} GPU did not experience any significant errors or inconsistencies.

In batch tests (BT), performance generally decreases as batch size increases across all devices, except for the \gls{avp} GPU, which remains stable regardless of batch size. The \gls{avp} CPU, however, experiences a strong performance degradation as batch size increases, more pronounced than in other devices. For thread tests (TT), across all four devices, mid-range thread counts (4 to 8) yielded the best performance, while lower (1, 2) and higher (16, 32) thread counts resulted in lower performance. Notably, the \gls{avp} CPU fails at thread counts of 16 and 32, whereas the \gls{avp} GPU exhibits its best performance at these higher thread counts.

In terms of memory consumption, significant variations were observed across different models. Generally, larger models consumed more memory. Interestingly, each of the five model series displayed distinct memory consumption trends. Some smaller models from one series consumed more memory than larger models from another series. For instance, $m_2$ (with size 1.36 GB) consumes less memory than $m_6$ (with size 1.32 GB), highlighting the model-dependent nature of memory consumption (see~\cref{fig:Memory_plot}).

Battery consumption did not show significant differences across devices, although the \gls{avp} exhibited slightly higher battery life loss, with an average battery depletion of up to $10.1\%$ (with GPU) and $12.6\%$ (with CPU) over 10 minutes. In contrast, the Vivo X100 Pro performed the best, with only a $2.5\%$ loss, while the Magic Leap 2 demonstrated good battery efficiency with an $8.5\%$ loss. The Meta Quest 3 consumed an average of $9.7\%$ over the same duration.

With 3D Pareto Optimality, we found \gls{avp} leading overall, \gls{ml2} demonstrating high consistency and stability, \gls{vivo} being slower but still consistent in performance, and \gls{mq3} ranking lowest across all aspects. Among the models, $m_3$, and $m_5$ performed particularly well, with smaller models generally exhibiting better results. In contrast, models from the LLaMA-2 and Mistral-7B series were the slowest. We have analyzed the performance of an on-device LLM using 57 queries from an interactive application dataset. Our findings suggest that with targeted training and fine-tuning, the performance can be further improved. \\
To make the per-model throughput differences clearer, we summarize several key trends from the full PP/TG results in Supplementary Table~4. Smaller models such as Qwen-0.5B ($m_1$) and Vikhr-2B ($m_2$) are consistently the fastest across all devices, for example, on AVP-CPU, $m_2$ reaches up to 20.25~t/s in PP tests, whereas the largest model ($m_{17}$) achieves only 4.30~t/s, a 4.7$\times$ slowdown. Longer prompts further widen these gaps: on ML2, $m_5$ decreases from 15.59~t/s (64 tokens) to 9.34~t/s (1024 tokens), while on MQ3, $m_{11}$ drops from 8.73~t/s to 4.54~t/s. These observations clarify how model size and prompt length jointly influence throughput and provide quantitative guidance for selecting appropriate models for XR deployment.\\
These observations highlight that performance variance in XR environments is shaped not only by model size but also by device-level thermal and memory constraints. Such insights help developers anticipate instability and choose model-device pairs that maintain predictable behavior during real-time XR interaction.

 \subsection{The Impact of LLM Characteristics}
The size of the model on disk, which correlates with the number of parameters, has a substantial impact on performance. In this study, we primarily selected models below 4 GB in size to ensure they could be deployed on all tested devices. We observed that larger models not only exhibited lower inference speeds and higher memory usage but also consumed more battery compared to smaller ones. 
However, in the context of interactive application accuracy, larger models such as the Mistral-7B and Phi-3.1 series achieved higher accuracy, while LLaMA-2 performed comparatively poorly. It is important to note that our testing for interactive applications was conducted without any fine-tuning. This suggests that selecting a smaller model, for example, from the Qwen series; and fine-tuning it on XR-specific data could achieve a better balance of accuracy and efficiency. \\
Overall, these findings indicate the need to carefully balance model size and parameter count to reach optimal trade-offs between speed, memory consumption, battery life, and accuracy. For XR practitioners, these trends imply that model selection should be guided by the latency budget and interaction style of the target application, with smaller models preferred for continuous or reactive XR tasks and larger ones reserved for high-accuracy, non-time-critical scenarios.

 \subsection{The Impact of Device Capabilities}
We found that the \gls{avp} achieved the highest speed in both PP and TG tests. \gls{ml2} ranked second in overall performance, while the Vivo X100s Pro outperformed \gls{mq3} in TG performance but ranked fourth in PP speed. These results can be attributed to the AVP's unified memory architecture and hardware-accelerated memory management, which result in lower memory consumption and more stable performance. 
Furthermore, the combination of hardware design and operating system-level memory management plays a critical role in determining device efficiency. The remaining three devices demonstrated comparatively weaker performance, underscoring that selecting an XR device for AI workloads depends heavily on available memory and the underlying memory management capabilities. 
While understanding these factors beforehand is beneficial, applying \aivaluatexr provides a systematic way to quickly assess and compare device suitability for AI applications in XR environments.\\
This reinforces that hardware-software co-design is essential in XR AI systems: performance bottlenecks often arise from memory architecture and thermal management rather than raw compute. Our results offer guidance for selecting XR hardware aligned with intended AI workloads. These observations help developers align device choice with workload characteristics, for example, preferring AVP for GPU-accelerated tasks, ML2 for long-running stability, and avoiding MQ3 for workloads that require sustained operation under thermal stress.

\subsection{Implications for XR Applications}
Based on our evaluation of XR interactive applications, specifically VOICE and GeoVis (as shown in~\cref{tab:geovis_voice_grouped}), we observe that even without any domain-specific training, larger models achieved reasonably good accuracy. While smaller models demonstrated lower accuracy, their performance could likely be improved through grounding or fine-tuning on XR-specific datasets. The trade-offs between accuracy and processing time linked to model size are an important consideration for practical deployment but can be mitigated through targeted domain adaptation. 
These findings highlight the potential of \aivaluatexr as a practical tool for identifying and optimizing model-device configurations in this emerging area at the intersection of AI and XR.
These implications demonstrate that on-device LLMs are feasible for many XR use cases today, but their effectiveness depends strongly on aligning the model choice with the application's latency budget and interaction pattern.

\subsection{The Future of On-Device LLMs}
While cloud-based LLMs offer powerful capabilities, they may not always be preferred due to factors such as limited internet connectivity, subscription costs, privacy concerns, or regulatory restrictions. An alternative approach is the client-server model, where LLMs run on a local server and XR devices connect over a local network. However, this setup still introduces network latency (see~\cref{tab:latency7LLMs}) and does not fully address data security risks.
By contrast, running LLMs directly on-device makes the system more autonomous and independent of external infrastructure. Looking ahead, we anticipate that advances in model compression, quantization techniques, and specialized hardware accelerators will further reduce resource demands and latency, making fully on-device intelligence increasingly practical. Given the rapid growth in XR and wearable devices, enabling on-device LLMs has the potential to unlock a wide range of real-world applications across domains, from education and training to healthcare, industrial maintenance, and beyond. 
These trends indicate that on-device LLMs are poised to become increasingly central to future XR systems.

 \subsection{Insights for XR Practitioners}
\aivaluatexr provides a general methodology for deploying LLMs on XR devices and evaluating arbitrary LLMs across heterogeneous XR hardware. Based on our evaluation of the 68 model–device pairs, we distill the following actionable insights and recommendations to guide XR practitioners in deploying on-device LLMs:
\begin{itemize} 
\item The framework enables fully on-device LLM deployment, which is particularly relevant for XR applications requiring low latency, offline functionality, privacy preservation, and/or task-specific fine-tuning.
\item Smaller and mid-sized models (e.g., Qwen-0.5B and Vikhr-2B) provide an optimal trade-off among processing speed, memory usage, and energy efficiency for real-time interaction, while large size models are only practical when latency constraints are relaxed.
\item Moderate thread counts (4--8) consistently yield optimal performance on CPU-based XR devices, whereas higher thread counts often lead to thermal throttling and reduced throughput.
\item For interactive workloads, batch sizes should remain at or below 128 on CPU-based XR devices to avoid latency degradation; GPU-enabled platforms such as AVP remain stable across larger batch sizes.
\item Device characteristics significantly influence deployment choices: AVP-GPU achieves the highest throughput, Magic Leap~2 offers superior stability, Vivo excels in energy efficiency, and Meta Quest~3 shows higher variability under sustained workloads. 
\end{itemize} 
\subsection{Limitations}
Apart from \gls{avp}, our study focuses on CPU-based evaluation, as limited GPU memory on the three devices hindered full model loading, affecting performance.  
 Moreover, our evaluation centers on end-to-end performance, and therefore we do not present quantitative measurements of internal CPU/GPU utilization, which can vary across devices due to differences in thermal design, OS scheduling, and power-management policies.\\
 Another limitation is that our implementation relies on \textit{Llama.cpp}. We selected this framework because it offers broad support for GGUF models and consistent cross-device compatibility. While alternative runtimes such as ONNX Runtime or TensorFlow Lite may offer different performance characteristics, they do not yet support the full range of models, quantization formats, or device platforms considered in this study. Extending \textit{\aivaluatexr} to incorporate multiple inference backends is an important direction for future work.

A further limitation is that our experiments do not include long-duration evaluations (e.g., 30-60 minutes of continuous operation), which are essential for characterizing thermal throttling and stability during extended XR sessions. Battery consumption was estimated using device-reported percentage values rather than sensor-level power measurements, which may introduce small inaccuracies. In addition, our PP/TG experiments rely on synthetic prompts generated by \texttt{llama.cpp}; real XR usage may involve more diverse linguistic structures that manifest different latency behavior. Similarly, our interactive application tests were conducted on relatively constrained datasets, and real-world XR tasks may yield different accuracy profiles.

We also note that firmware versions, ambient temperature, and background OS activity can influence XR device performance, meaning that absolute timing values may vary across deployments. Finally, our interactive accuracy evaluation uses models without domain-specific fine-tuning; thus, the reported accuracy should be interpreted as a conservative baseline rather than an upper bound.

\section{Conclusion}
\label{sec:conclusion}
\textit{\aivaluatexr} provides a comprehensive framework for deploying LLMs on XR devices and conducting in-depth evaluations across multiple performance factors. Using this framework, we benchmarked 17 LLMs across four XR devices to assess their performance in a systematic and reproducible manner.
 Given the GPU memory limitations of most XR devices, our study primarily focuses on CPU-based analysis. However, since the \gls{avp} has sufficient GPU resources, we also include GPU results for \gls{avp}, providing additional insights into the study.
 Generally, for processing speed, we can rank the four devices in the following order: \gls{avp}, \gls{ml2}, \gls{vivo}, and \gls{mq3}, although \gls{mq3} outperforms \gls{vivo} in TG.
 
In memory consumption, \gls{avp} was found the most memory-efficient. Among the remaining devices, differences were minimal, though \gls{ml2} occasionally consumed the most memory. As expected, larger models required more memory, but this trend varied across different model series. For BT tests, a lower batch size (128) yielded the best performance, while larger batch sizes reduced speed for CPU-based tests, with \gls{avp} GPU remaining largely unaffected. For TT tests, medium thread counts (4--8) were optimal across all devices. In terms of consistency, \gls{ml2} is the most stable, exhibiting the lowest variance in speed and zero errors. \gls{avp} GPU also maintains low variance. \gls{mq3} performs the worst in this aspect, while \gls{vivo} is more stable than \gls{mq3} but less consistent than the rest. 
The Pareto analysis highlighted several models on \gls{avp} due to their outstanding performance, while the Pareto fronts found on \gls{ml2} because of their consistently stable results. Additionally, a few models on \gls{vivo} and \gls{mq3} emerged on the Pareto front because of their balanced trade-offs between performance and stability. Models $m_{12}$ to $m_{17}$ were found slower.

Our latency comparison across cloud-based, client-server, and on-device deployments—along with two interactive application tests—demonstrates that while on-device LLMs currently yield slightly lower accuracy, this can be improved through domain-specific fine-tuning. Notably, despite longer warm-up times (\ie higher \textit{Cold F.R.T}), on-device models offer excellent latency post-loading (\ie lower \textit{Warm F.R.T}), making them highly promising for real-time XR applications. Given the rise of XR and LLMs, we believe on-device AI has strong potential, in which \aivaluatexr\ can play a key role.

%% file: biography.tex
\vskip -0.1279996478cm 
\begin{IEEEbiography}[{\includegraphics[width=1in,height=1.25in,clip,keepaspectratio]{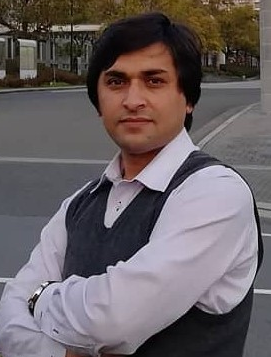}}]{Dawar~Khan}
received the Ph.D. degree from NLPR, Institute of Automation, UCAS, Beijing, China, in 2018. 
He was an Assistant Professor at NAIST, Japan (2018–2020), and at the University of Haripur (UOH), Pakistan, until 2022. 
From 2022 to 2025, he was a Postdoctoral Fellow at KAUST, Saudi Arabia. 
Since 2025, he has been an Assistant Professor at UOH. 
His research interests include computer graphics, computer vision, extended reality, HCI, and AI for 3D scene understanding.
\end{IEEEbiography}
    \vskip -1.09996478cm 
    \begin{IEEEbiography}[{\includegraphics[width=1in,height=1.25in,clip,keepaspectratio]{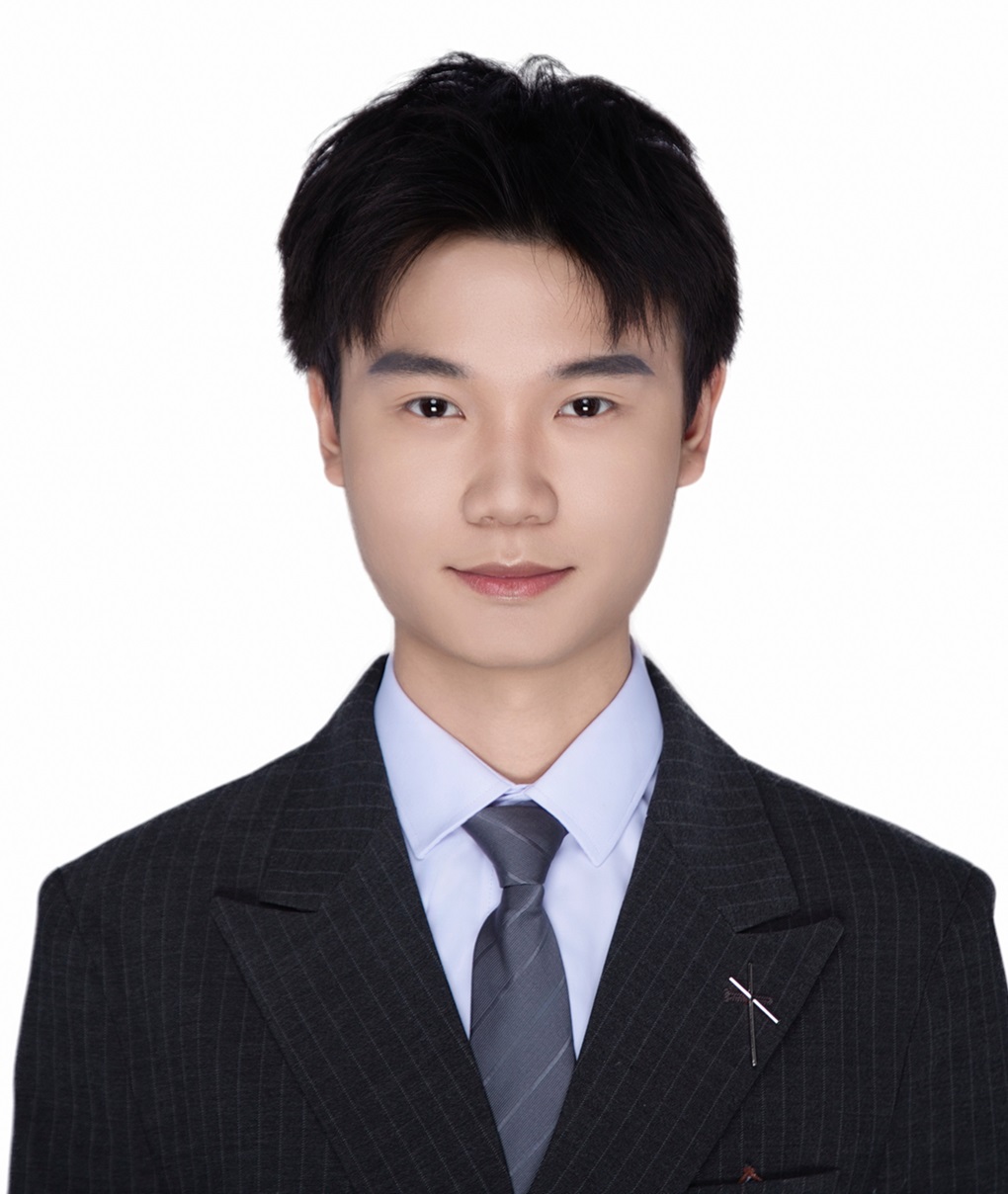}}]{Xinyu~Liu}
is a Master's student at  KAUST. 
He received his B.Eng. degree in Internet of Things Engineering from the University of Electronic Science and Technology of China (UESTC), China. 
In 2024, he participated in the Visiting Student Research Program (VSRP) at KAUST. 
His research interests include IoT, extended reality (XR), and large language models.
\end{IEEEbiography}
    \vskip -1.09996478cm 
\begin{IEEEbiography}[{\includegraphics[width=1in,height=1.25in,clip,keepaspectratio]{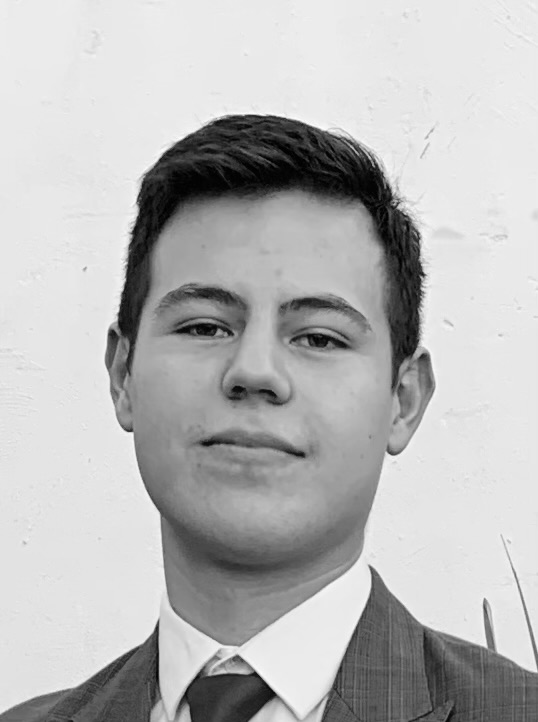}}]{Omar~Mena}
 is a PhD student in Computer Science under the supervision of Professor Ivan Viola at  KAUST. He obtained his BSc in Artificial Intelligence at Universidad Panamericana, Mexico, and his MSc in Computer Science from KAUST. His research interests are XR human-computer interaction and computer graphics.
\end{IEEEbiography}
    \vskip -1.09996478cm 
\begin{IEEEbiography}[{\includegraphics[width=1in,height=1.25in,clip,keepaspectratio]{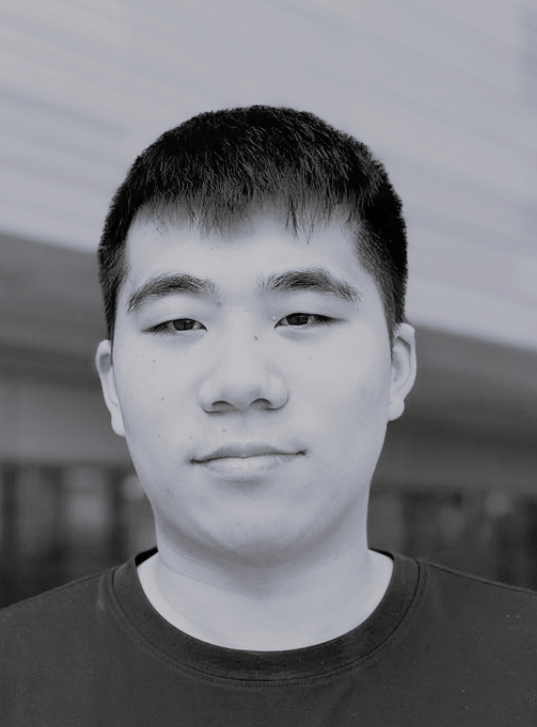}}]{Donggang~Jia}
 is a Computer Science PhD student at KAUST, where he is part of the nanovisualization group. He received a B. Eng degree in Computer Science from the Beijing University of Posts and Telecommunications, China, and an MSc in Artificial Intelligence from the University of Southampton, UK. His research interests include scientific visualization, human-computer interaction, and computer graphics.
\end{IEEEbiography}
    \vskip -1.09996478cm 
\begin{IEEEbiography}[{\includegraphics[width=1in,height=1.25in,clip,keepaspectratio]{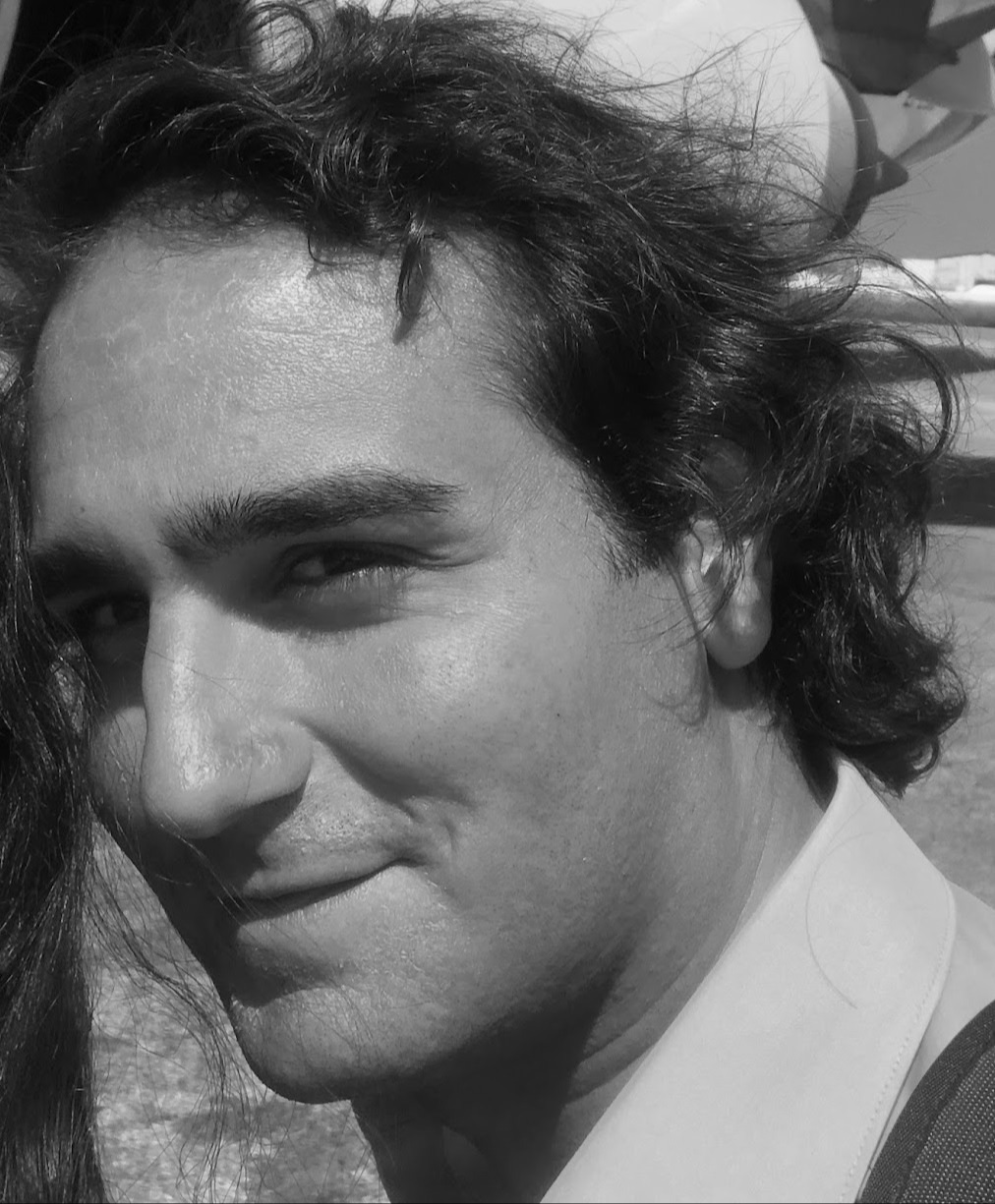}}]{Alexandre~Kouyoumdjian}
is a research scientist at King Abdullah University of Science and Technology (KAUST). He holds a PhD in computer science from University Paris-Saclay. He conducts research on multiscale visualization, interaction, and modeling for biology, with an additional focus on virtual and augmented reality.
\end{IEEEbiography}
    \vskip -1.09996478cm 
\begin{IEEEbiography}[{\includegraphics[width=1in,height=1.25in,clip,keepaspectratio]{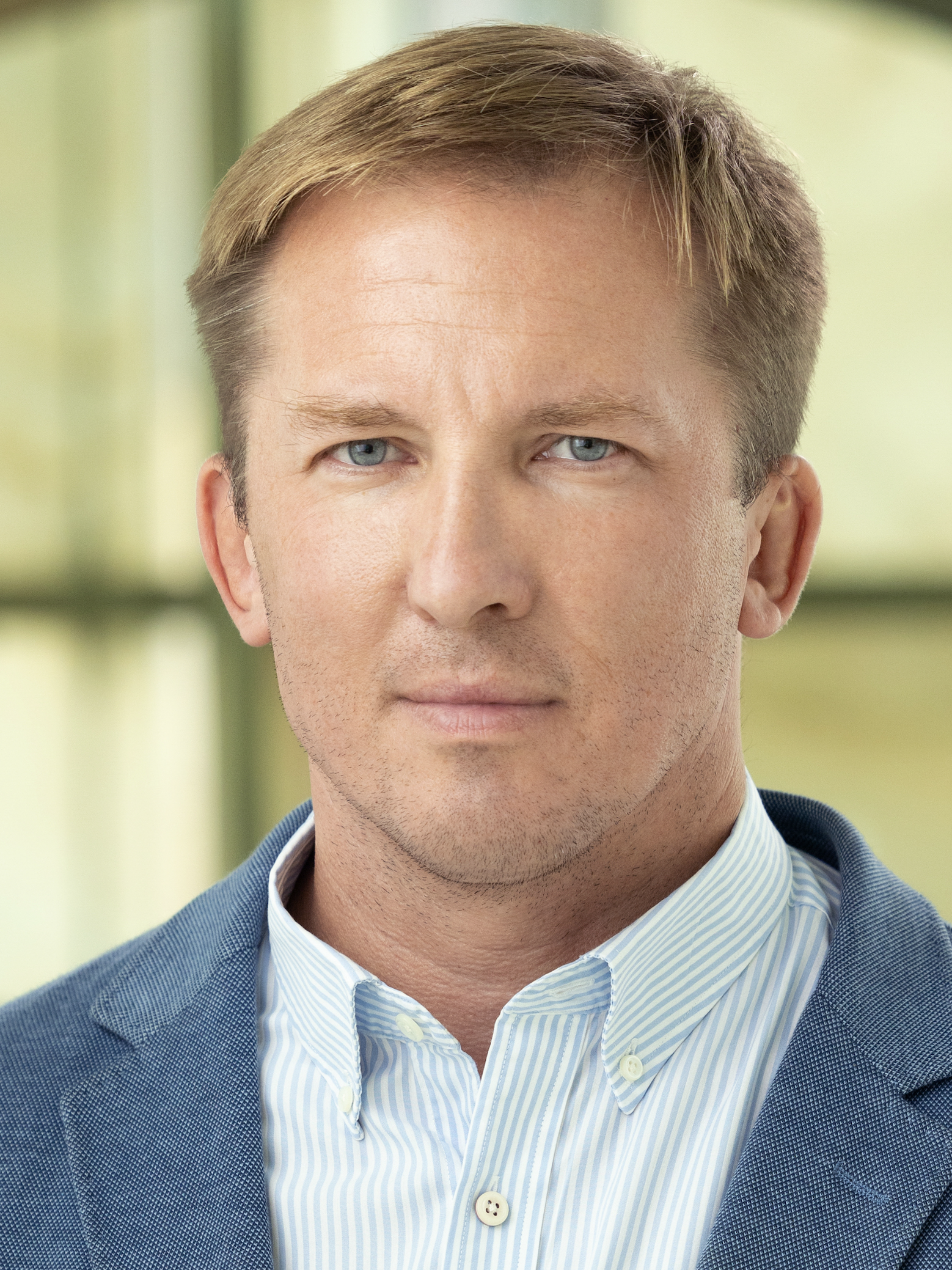}}]{Ivan~Viola}
is a professor at KAUST, Saudi Arabia. He graduated from TU Wien, Austria. In 2005 he took a postdoc position at the University of Bergen, Norway, where he was gradually promoted to the professor rank. In 2013 he received a WWTF grant to establish a research group at TU Wien. At KAUST, he continues developing new visualization techniques, primarily oriented on data reconstruction, interpretation, representation, modeling, and rendering.
\end{IEEEbiography}